\documentclass[longauth]{aaEC}

\usepackage{graphicx}
\usepackage{natbib}
\usepackage{scalerel}
\usepackage{multirow}

\usepackage[table]{xcolor}

\usepackage{txfonts}
\usepackage[pdfencoding=auto,psdextra]{hyperref}
\hypersetup{
    colorlinks=true,
    linkcolor=blue,
    filecolor=magenta,      
    urlcolor=blue,
    citecolor=blue
}
\makeatletter
\renewcommand*\aa@pageof{, page \thepage{} of \pageref*{LastPage}}
\makeatother

\newcommand\T{\rule{0pt}{2.6ex}}     
\newcommand\B{\rule[-1.2ex]{0pt}{0pt}}

\newcommand{\spitzer}{\textit{Spitzer}\xspace}
\newcommand{\sextractor}{\texttt{SourceExtractor}\xspace}
\newcommand{\tphot}{\texttt{t-phot}\xspace}
\newcommand{\ihsc}{\ensuremath{i_\sfont{HSC}}\xspace}

\usepackage[utf8]{inputenc}

\usepackage[switch, modulo]{lineno}

\usepackage{euclid}

\usepackage{cleveref}
\crefname{section}{Sect.}{Sects.}
\Crefname{section}{Section}{Sections}
\crefname{figure}{Fig.}{Figs.}
\Crefname{figure}{Figure}{Figures}
\crefname{table}{Table}{Tables}
\Crefname{table}{Table}{Tables}
\crefname{appendix}{Appendix}{Appendices}
\Crefname{appendix}{Appendix}{Appendices}

\usepackage{glossaries}

\newacronym{q1}{Q1}{Euclid Quick Data Release 1}
\newacronym{eds}{EDS}{Euclid Deep Survey}
\newacronym{ews}{EWS}{Euclid Wide Survey}
\newacronym{edf}{EDFs}{Euclid Deep Fields}
\newacronym{edfn}{EDF-N}{Euclid Deep Field North}
\newacronym{edfs}{EDF-S}{Euclid Deep Field South}
\newacronym{edff}{EDF-F}{Euclid Deep Field Fornax}
\newacronym{cdfs}{CDF-S}{Chandra Deep Field South}
\newacronym{nir}{NIR}{Near Infrared}
\newacronym{vis}{VIS}{Visible Instrument}
\newacronym{nisp}{NISP}{Near Infrared Spectrometer and Photometer}
\newacronym{ou}{OU}{Organisational Unit}
\newacronym{sgs}{SGS}{Science Ground Segment}
\newacronym{hsc}{HSC}{Hyper-Suprime Cam}
\newacronym{h20}{H20}{Hawaii Twenty Square Degree Survey}
\newacronym{gaea}{GAEA}{GAlaxy Evolution and Assembly}
\newacronym{psf}{PSF}{point-spread function}
\newacronym{agn}{AGN}{active galactic nuclei}
\newacronym{hst}{HST}{\textit{Hubble} Space Telescope}
\newacronym{wfc3}{WFC3}{Wide Field Camera 3}
\newacronym{irac}{IRAC}{\spitzer Infrared Array Camera}
\newacronym{dawn}{DAWN}{Cosmic Dawn Survey}
\newacronym{spuds}{SpUDS}{United Kingdom Infrared Telescope Infrared Deep Sky Survey Ultra-Deep Survey}
\newacronym{rlqs}{RLQs}{radio-loud quasars}
\newacronym{hzrgs}{HzRGs}{high-redshift radio galaxies}
\newacronym{efeds}{eFEDS}{eROSITA Equatorial Depth Survey}
\newacronym{erosita}{eRASS1}{SRG/eROSITA All-Sky Survey}
\newacronym{snr}{S/N}{signal-to-noise ratio}

\begin{document}
%
%

\title{Euclid Quick Data Release (Q1)} \subtitle{Combined \Euclid and \spitzer galaxy density catalogues at $z$>1.3 and detection of significant \Euclid passive galaxy overdensities in \spitzer overdense regions}



\newcommand{\orcid}[1]{} 
\newcommand{\cressida}[1]{\textcolor{magenta}{#1}}

\renewcommand{\orcid}[1]{} 


\author{Euclid Collaboration: N.~Mai\inst{\ref{aff1},\ref{aff2}}
\and S.~Mei\orcid{0000-0002-2849-559X}\thanks{\email{mei@apc.in2p3.fr}}\inst{\ref{aff1},\ref{aff2}}
\and C.~Cleland\inst{\ref{aff1}}
\and R.~Chary\orcid{0000-0001-7583-0621}\inst{\ref{aff3},\ref{aff4}}
\and J.~G.~Bartlett\orcid{0000-0002-0685-8310}\inst{\ref{aff1}}
\and G.~Castignani\orcid{0000-0001-6831-0687}\inst{\ref{aff5}}
\and H.~Dannerbauer\orcid{0000-0001-7147-3575}\inst{\ref{aff6}}
\and G.~De~Lucia\orcid{0000-0002-6220-9104}\inst{\ref{aff7}}
\and F.~Fontanot\orcid{0000-0003-4744-0188}\inst{\ref{aff7},\ref{aff8}}
\and D.~Scott\orcid{0000-0002-6878-9840}\inst{\ref{aff9}}
\and S.~Andreon\orcid{0000-0002-2041-8784}\inst{\ref{aff10}}
\and S.~Bhargava\orcid{0000-0003-3851-7219}\inst{\ref{aff11}}
\and H.~Dole\orcid{0000-0002-9767-3839}\inst{\ref{aff12}}
\and T.~Dussere\inst{\ref{aff12}}
\and S.~A.~Stanford\orcid{0000-0003-0122-0841}\inst{\ref{aff13}}
\and V.~P.~Tran\inst{\ref{aff1},\ref{aff2}}
\and J.~R.~Weaver\orcid{0000-0003-1614-196X}\inst{\ref{aff14}}
\and P.-A.~Duc\orcid{0000-0003-3343-6284}\inst{\ref{aff15}}
\and I.~Risso\orcid{0000-0003-2525-7761}\inst{\ref{aff16}}
\and N.~Aghanim\orcid{0000-0002-6688-8992}\inst{\ref{aff12}}
\and B.~Altieri\orcid{0000-0003-3936-0284}\inst{\ref{aff17}}
\and A.~Amara\inst{\ref{aff18}}
\and N.~Auricchio\orcid{0000-0003-4444-8651}\inst{\ref{aff5}}
\and H.~Aussel\orcid{0000-0002-1371-5705}\inst{\ref{aff19}}
\and C.~Baccigalupi\orcid{0000-0002-8211-1630}\inst{\ref{aff8},\ref{aff7},\ref{aff20},\ref{aff21}}
\and M.~Baldi\orcid{0000-0003-4145-1943}\inst{\ref{aff22},\ref{aff5},\ref{aff23}}
\and A.~Balestra\orcid{0000-0002-6967-261X}\inst{\ref{aff24}}
\and S.~Bardelli\orcid{0000-0002-8900-0298}\inst{\ref{aff5}}
\and P.~Battaglia\orcid{0000-0002-7337-5909}\inst{\ref{aff5}}
\and A.~Biviano\orcid{0000-0002-0857-0732}\inst{\ref{aff7},\ref{aff8}}
\and A.~Bonchi\orcid{0000-0002-2667-5482}\inst{\ref{aff25}}
\and D.~Bonino\orcid{0000-0002-3336-9977}\inst{\ref{aff26}}
\and E.~Branchini\orcid{0000-0002-0808-6908}\inst{\ref{aff27},\ref{aff28},\ref{aff10}}
\and M.~Brescia\orcid{0000-0001-9506-5680}\inst{\ref{aff29},\ref{aff30}}
\and J.~Brinchmann\orcid{0000-0003-4359-8797}\inst{\ref{aff31},\ref{aff32}}
\and A.~Caillat\inst{\ref{aff33}}
\and S.~Camera\orcid{0000-0003-3399-3574}\inst{\ref{aff34},\ref{aff35},\ref{aff26}}
\and G.~Ca\~nas-Herrera\orcid{0000-0003-2796-2149}\inst{\ref{aff36},\ref{aff37},\ref{aff38}}
\and V.~Capobianco\orcid{0000-0002-3309-7692}\inst{\ref{aff26}}
\and C.~Carbone\orcid{0000-0003-0125-3563}\inst{\ref{aff39}}
\and J.~Carretero\orcid{0000-0002-3130-0204}\inst{\ref{aff40},\ref{aff41}}
\and S.~Casas\orcid{0000-0002-4751-5138}\inst{\ref{aff42}}
\and M.~Castellano\orcid{0000-0001-9875-8263}\inst{\ref{aff43}}
\and S.~Cavuoti\orcid{0000-0002-3787-4196}\inst{\ref{aff30},\ref{aff44}}
\and K.~C.~Chambers\orcid{0000-0001-6965-7789}\inst{\ref{aff45}}
\and A.~Cimatti\inst{\ref{aff46}}
\and C.~Colodro-Conde\inst{\ref{aff47}}
\and G.~Congedo\orcid{0000-0003-2508-0046}\inst{\ref{aff48}}
\and C.~J.~Conselice\orcid{0000-0003-1949-7638}\inst{\ref{aff49}}
\and L.~Conversi\orcid{0000-0002-6710-8476}\inst{\ref{aff50},\ref{aff17}}
\and Y.~Copin\orcid{0000-0002-5317-7518}\inst{\ref{aff51}}
\and F.~Courbin\orcid{0000-0003-0758-6510}\inst{\ref{aff52},\ref{aff53}}
\and H.~M.~Courtois\orcid{0000-0003-0509-1776}\inst{\ref{aff54}}
\and A.~Da~Silva\orcid{0000-0002-6385-1609}\inst{\ref{aff55},\ref{aff56}}
\and H.~Degaudenzi\orcid{0000-0002-5887-6799}\inst{\ref{aff57}}
\and A.~M.~Di~Giorgio\orcid{0000-0002-4767-2360}\inst{\ref{aff58}}
\and M.~Douspis\orcid{0000-0003-4203-3954}\inst{\ref{aff12}}
\and F.~Dubath\orcid{0000-0002-6533-2810}\inst{\ref{aff57}}
\and C.~A.~J.~Duncan\orcid{0009-0003-3573-0791}\inst{\ref{aff49}}
\and X.~Dupac\inst{\ref{aff17}}
\and S.~Dusini\orcid{0000-0002-1128-0664}\inst{\ref{aff59}}
\and A.~Ealet\orcid{0000-0003-3070-014X}\inst{\ref{aff51}}
\and S.~Escoffier\orcid{0000-0002-2847-7498}\inst{\ref{aff60}}
\and M.~Farina\orcid{0000-0002-3089-7846}\inst{\ref{aff58}}
\and R.~Farinelli\inst{\ref{aff5}}
\and F.~Faustini\orcid{0000-0001-6274-5145}\inst{\ref{aff43},\ref{aff25}}
\and S.~Ferriol\inst{\ref{aff51}}
\and F.~Finelli\orcid{0000-0002-6694-3269}\inst{\ref{aff5},\ref{aff61}}
\and P.~Fosalba\orcid{0000-0002-1510-5214}\inst{\ref{aff62},\ref{aff63}}
\and S.~Fotopoulou\orcid{0000-0002-9686-254X}\inst{\ref{aff64}}
\and M.~Frailis\orcid{0000-0002-7400-2135}\inst{\ref{aff7}}
\and E.~Franceschi\orcid{0000-0002-0585-6591}\inst{\ref{aff5}}
\and M.~Fumana\orcid{0000-0001-6787-5950}\inst{\ref{aff39}}
\and S.~Galeotta\orcid{0000-0002-3748-5115}\inst{\ref{aff7}}
\and K.~George\orcid{0000-0002-1734-8455}\inst{\ref{aff65}}
\and B.~Gillis\orcid{0000-0002-4478-1270}\inst{\ref{aff48}}
\and C.~Giocoli\orcid{0000-0002-9590-7961}\inst{\ref{aff5},\ref{aff23}}
\and J.~Gracia-Carpio\inst{\ref{aff66}}
\and B.~R.~Granett\orcid{0000-0003-2694-9284}\inst{\ref{aff10}}
\and A.~Grazian\orcid{0000-0002-5688-0663}\inst{\ref{aff24}}
\and F.~Grupp\inst{\ref{aff66},\ref{aff65}}
\and S.~V.~H.~Haugan\orcid{0000-0001-9648-7260}\inst{\ref{aff67}}
\and W.~Holmes\inst{\ref{aff68}}
\and F.~Hormuth\inst{\ref{aff69}}
\and A.~Hornstrup\orcid{0000-0002-3363-0936}\inst{\ref{aff70},\ref{aff71}}
\and P.~Hudelot\inst{\ref{aff72}}
\and K.~Jahnke\orcid{0000-0003-3804-2137}\inst{\ref{aff73}}
\and M.~Jhabvala\inst{\ref{aff74}}
\and E.~Keih\"anen\orcid{0000-0003-1804-7715}\inst{\ref{aff75}}
\and S.~Kermiche\orcid{0000-0002-0302-5735}\inst{\ref{aff60}}
\and A.~Kiessling\orcid{0000-0002-2590-1273}\inst{\ref{aff68}}
\and B.~Kubik\orcid{0009-0006-5823-4880}\inst{\ref{aff51}}
\and M.~K\"ummel\orcid{0000-0003-2791-2117}\inst{\ref{aff65}}
\and M.~Kunz\orcid{0000-0002-3052-7394}\inst{\ref{aff76}}
\and H.~Kurki-Suonio\orcid{0000-0002-4618-3063}\inst{\ref{aff77},\ref{aff78}}
\and Q.~Le~Boulc'h\inst{\ref{aff79}}
\and A.~M.~C.~Le~Brun\orcid{0000-0002-0936-4594}\inst{\ref{aff80}}
\and D.~Le~Mignant\orcid{0000-0002-5339-5515}\inst{\ref{aff33}}
\and S.~Ligori\orcid{0000-0003-4172-4606}\inst{\ref{aff26}}
\and P.~B.~Lilje\orcid{0000-0003-4324-7794}\inst{\ref{aff67}}
\and V.~Lindholm\orcid{0000-0003-2317-5471}\inst{\ref{aff77},\ref{aff78}}
\and I.~Lloro\orcid{0000-0001-5966-1434}\inst{\ref{aff81}}
\and G.~Mainetti\orcid{0000-0003-2384-2377}\inst{\ref{aff79}}
\and D.~Maino\inst{\ref{aff82},\ref{aff39},\ref{aff83}}
\and E.~Maiorano\orcid{0000-0003-2593-4355}\inst{\ref{aff5}}
\and O.~Mansutti\orcid{0000-0001-5758-4658}\inst{\ref{aff7}}
\and O.~Marggraf\orcid{0000-0001-7242-3852}\inst{\ref{aff84}}
\and M.~Martinelli\orcid{0000-0002-6943-7732}\inst{\ref{aff43},\ref{aff85}}
\and N.~Martinet\orcid{0000-0003-2786-7790}\inst{\ref{aff33}}
\and F.~Marulli\orcid{0000-0002-8850-0303}\inst{\ref{aff86},\ref{aff5},\ref{aff23}}
\and R.~Massey\orcid{0000-0002-6085-3780}\inst{\ref{aff87}}
\and S.~Maurogordato\inst{\ref{aff11}}
\and E.~Medinaceli\orcid{0000-0002-4040-7783}\inst{\ref{aff5}}
\and Y.~Mellier\inst{\ref{aff88},\ref{aff72}}
\and M.~Meneghetti\orcid{0000-0003-1225-7084}\inst{\ref{aff5},\ref{aff23}}
\and E.~Merlin\orcid{0000-0001-6870-8900}\inst{\ref{aff43}}
\and G.~Meylan\inst{\ref{aff89}}
\and A.~Mora\orcid{0000-0002-1922-8529}\inst{\ref{aff90}}
\and M.~Moresco\orcid{0000-0002-7616-7136}\inst{\ref{aff86},\ref{aff5}}
\and L.~Moscardini\orcid{0000-0002-3473-6716}\inst{\ref{aff86},\ref{aff5},\ref{aff23}}
\and R.~Nakajima\orcid{0009-0009-1213-7040}\inst{\ref{aff84}}
\and C.~Neissner\orcid{0000-0001-8524-4968}\inst{\ref{aff91},\ref{aff41}}
\and S.-M.~Niemi\inst{\ref{aff36}}
\and J.~W.~Nightingale\orcid{0000-0002-8987-7401}\inst{\ref{aff92}}
\and C.~Padilla\orcid{0000-0001-7951-0166}\inst{\ref{aff91}}
\and S.~Paltani\orcid{0000-0002-8108-9179}\inst{\ref{aff57}}
\and F.~Pasian\orcid{0000-0002-4869-3227}\inst{\ref{aff7}}
\and K.~Pedersen\inst{\ref{aff93}}
\and W.~J.~Percival\orcid{0000-0002-0644-5727}\inst{\ref{aff94},\ref{aff95},\ref{aff96}}
\and V.~Pettorino\inst{\ref{aff36}}
\and S.~Pires\orcid{0000-0002-0249-2104}\inst{\ref{aff19}}
\and G.~Polenta\orcid{0000-0003-4067-9196}\inst{\ref{aff25}}
\and M.~Poncet\inst{\ref{aff97}}
\and L.~A.~Popa\inst{\ref{aff98}}
\and L.~Pozzetti\orcid{0000-0001-7085-0412}\inst{\ref{aff5}}
\and F.~Raison\orcid{0000-0002-7819-6918}\inst{\ref{aff66}}
\and A.~Renzi\orcid{0000-0001-9856-1970}\inst{\ref{aff99},\ref{aff59}}
\and J.~Rhodes\orcid{0000-0002-4485-8549}\inst{\ref{aff68}}
\and G.~Riccio\inst{\ref{aff30}}
\and E.~Romelli\orcid{0000-0003-3069-9222}\inst{\ref{aff7}}
\and M.~Roncarelli\orcid{0000-0001-9587-7822}\inst{\ref{aff5}}
\and R.~Saglia\orcid{0000-0003-0378-7032}\inst{\ref{aff65},\ref{aff66}}
\and Z.~Sakr\orcid{0000-0002-4823-3757}\inst{\ref{aff100},\ref{aff101},\ref{aff102}}
\and D.~Sapone\orcid{0000-0001-7089-4503}\inst{\ref{aff103}}
\and B.~Sartoris\orcid{0000-0003-1337-5269}\inst{\ref{aff65},\ref{aff7}}
\and J.~A.~Schewtschenko\orcid{0000-0002-4913-6393}\inst{\ref{aff48}}
\and P.~Schneider\orcid{0000-0001-8561-2679}\inst{\ref{aff84}}
\and M.~Scodeggio\inst{\ref{aff39}}
\and A.~Secroun\orcid{0000-0003-0505-3710}\inst{\ref{aff60}}
\and G.~Seidel\orcid{0000-0003-2907-353X}\inst{\ref{aff73}}
\and M.~Seiffert\orcid{0000-0002-7536-9393}\inst{\ref{aff68}}
\and S.~Serrano\orcid{0000-0002-0211-2861}\inst{\ref{aff62},\ref{aff104},\ref{aff63}}
\and P.~Simon\inst{\ref{aff84}}
\and C.~Sirignano\orcid{0000-0002-0995-7146}\inst{\ref{aff99},\ref{aff59}}
\and G.~Sirri\orcid{0000-0003-2626-2853}\inst{\ref{aff23}}
\and L.~Stanco\orcid{0000-0002-9706-5104}\inst{\ref{aff59}}
\and J.~Steinwagner\orcid{0000-0001-7443-1047}\inst{\ref{aff66}}
\and P.~Tallada-Cresp\'{i}\orcid{0000-0002-1336-8328}\inst{\ref{aff40},\ref{aff41}}
\and A.~N.~Taylor\inst{\ref{aff48}}
\and H.~I.~Teplitz\orcid{0000-0002-7064-5424}\inst{\ref{aff3}}
\and I.~Tereno\inst{\ref{aff55},\ref{aff105}}
\and R.~Toledo-Moreo\orcid{0000-0002-2997-4859}\inst{\ref{aff106}}
\and F.~Torradeflot\orcid{0000-0003-1160-1517}\inst{\ref{aff41},\ref{aff40}}
\and I.~Tutusaus\orcid{0000-0002-3199-0399}\inst{\ref{aff101}}
\and L.~Valenziano\orcid{0000-0002-1170-0104}\inst{\ref{aff5},\ref{aff61}}
\and J.~Valiviita\orcid{0000-0001-6225-3693}\inst{\ref{aff77},\ref{aff78}}
\and T.~Vassallo\orcid{0000-0001-6512-6358}\inst{\ref{aff65},\ref{aff7}}
\and G.~Verdoes~Kleijn\orcid{0000-0001-5803-2580}\inst{\ref{aff107}}
\and A.~Veropalumbo\orcid{0000-0003-2387-1194}\inst{\ref{aff10},\ref{aff28},\ref{aff27}}
\and Y.~Wang\orcid{0000-0002-4749-2984}\inst{\ref{aff3}}
\and J.~Weller\orcid{0000-0002-8282-2010}\inst{\ref{aff65},\ref{aff66}}
\and A.~Zacchei\orcid{0000-0003-0396-1192}\inst{\ref{aff7},\ref{aff8}}
\and G.~Zamorani\orcid{0000-0002-2318-301X}\inst{\ref{aff5}}
\and F.~M.~Zerbi\inst{\ref{aff10}}
\and E.~Zucca\orcid{0000-0002-5845-8132}\inst{\ref{aff5}}
\and V.~Allevato\orcid{0000-0001-7232-5152}\inst{\ref{aff30}}
\and M.~Ballardini\orcid{0000-0003-4481-3559}\inst{\ref{aff108},\ref{aff109},\ref{aff5}}
\and M.~Bolzonella\orcid{0000-0003-3278-4607}\inst{\ref{aff5}}
\and E.~Bozzo\orcid{0000-0002-8201-1525}\inst{\ref{aff57}}
\and C.~Burigana\orcid{0000-0002-3005-5796}\inst{\ref{aff110},\ref{aff61}}
\and R.~Cabanac\orcid{0000-0001-6679-2600}\inst{\ref{aff101}}
\and A.~Cappi\inst{\ref{aff5},\ref{aff11}}
\and D.~Di~Ferdinando\inst{\ref{aff23}}
\and J.~A.~Escartin~Vigo\inst{\ref{aff66}}
\and L.~Gabarra\orcid{0000-0002-8486-8856}\inst{\ref{aff111}}
\and W.~G.~Hartley\inst{\ref{aff57}}
\and J.~Mart\'{i}n-Fleitas\orcid{0000-0002-8594-569X}\inst{\ref{aff90}}
\and S.~Matthew\orcid{0000-0001-8448-1697}\inst{\ref{aff48}}
\and N.~Mauri\orcid{0000-0001-8196-1548}\inst{\ref{aff46},\ref{aff23}}
\and R.~B.~Metcalf\orcid{0000-0003-3167-2574}\inst{\ref{aff86},\ref{aff5}}
\and A.~Pezzotta\orcid{0000-0003-0726-2268}\inst{\ref{aff112},\ref{aff66}}
\and M.~P\"ontinen\orcid{0000-0001-5442-2530}\inst{\ref{aff77}}
\and C.~Porciani\orcid{0000-0002-7797-2508}\inst{\ref{aff84}}
\and V.~Scottez\inst{\ref{aff88},\ref{aff113}}
\and M.~Sereno\orcid{0000-0003-0302-0325}\inst{\ref{aff5},\ref{aff23}}
\and M.~Tenti\orcid{0000-0002-4254-5901}\inst{\ref{aff23}}
\and M.~Viel\orcid{0000-0002-2642-5707}\inst{\ref{aff8},\ref{aff7},\ref{aff21},\ref{aff20},\ref{aff114}}
\and M.~Wiesmann\orcid{0009-0000-8199-5860}\inst{\ref{aff67}}
\and Y.~Akrami\orcid{0000-0002-2407-7956}\inst{\ref{aff115},\ref{aff116}}
\and S.~Alvi\orcid{0000-0001-5779-8568}\inst{\ref{aff108}}
\and I.~T.~Andika\orcid{0000-0001-6102-9526}\inst{\ref{aff117},\ref{aff118}}
\and S.~Anselmi\orcid{0000-0002-3579-9583}\inst{\ref{aff59},\ref{aff99},\ref{aff119}}
\and M.~Archidiacono\orcid{0000-0003-4952-9012}\inst{\ref{aff82},\ref{aff83}}
\and F.~Atrio-Barandela\orcid{0000-0002-2130-2513}\inst{\ref{aff120}}
\and D.~Bertacca\orcid{0000-0002-2490-7139}\inst{\ref{aff99},\ref{aff24},\ref{aff59}}
\and M.~Bethermin\orcid{0000-0002-3915-2015}\inst{\ref{aff15}}
\and A.~Blanchard\orcid{0000-0001-8555-9003}\inst{\ref{aff101}}
\and L.~Blot\orcid{0000-0002-9622-7167}\inst{\ref{aff121},\ref{aff80}}
\and H.~B\"ohringer\orcid{0000-0001-8241-4204}\inst{\ref{aff66},\ref{aff122},\ref{aff123}}
\and S.~Borgani\orcid{0000-0001-6151-6439}\inst{\ref{aff124},\ref{aff8},\ref{aff7},\ref{aff20},\ref{aff114}}
\and M.~L.~Brown\orcid{0000-0002-0370-8077}\inst{\ref{aff49}}
\and S.~Bruton\orcid{0000-0002-6503-5218}\inst{\ref{aff125}}
\and A.~Calabro\orcid{0000-0003-2536-1614}\inst{\ref{aff43}}
\and F.~Caro\inst{\ref{aff43}}
\and C.~S.~Carvalho\inst{\ref{aff105}}
\and T.~Castro\orcid{0000-0002-6292-3228}\inst{\ref{aff7},\ref{aff20},\ref{aff8},\ref{aff114}}
\and Y.~Charles\inst{\ref{aff33}}
\and F.~Cogato\orcid{0000-0003-4632-6113}\inst{\ref{aff86},\ref{aff5}}
\and A.~R.~Cooray\orcid{0000-0002-3892-0190}\inst{\ref{aff126}}
\and O.~Cucciati\orcid{0000-0002-9336-7551}\inst{\ref{aff5}}
\and S.~Davini\orcid{0000-0003-3269-1718}\inst{\ref{aff28}}
\and F.~De~Paolis\orcid{0000-0001-6460-7563}\inst{\ref{aff127},\ref{aff128},\ref{aff129}}
\and G.~Desprez\orcid{0000-0001-8325-1742}\inst{\ref{aff107}}
\and A.~D\'iaz-S\'anchez\orcid{0000-0003-0748-4768}\inst{\ref{aff130}}
\and J.~J.~Diaz\inst{\ref{aff6},\ref{aff47}}
\and S.~Di~Domizio\orcid{0000-0003-2863-5895}\inst{\ref{aff27},\ref{aff28}}
\and J.~M.~Diego\orcid{0000-0001-9065-3926}\inst{\ref{aff131}}
\and P.~Dimauro\orcid{0000-0001-7399-2854}\inst{\ref{aff43},\ref{aff132}}
\and A.~Enia\orcid{0000-0002-0200-2857}\inst{\ref{aff22},\ref{aff5}}
\and Y.~Fang\inst{\ref{aff65}}
\and A.~G.~Ferrari\orcid{0009-0005-5266-4110}\inst{\ref{aff23}}
\and A.~Finoguenov\orcid{0000-0002-4606-5403}\inst{\ref{aff77}}
\and A.~Fontana\orcid{0000-0003-3820-2823}\inst{\ref{aff43}}
\and A.~Franco\orcid{0000-0002-4761-366X}\inst{\ref{aff128},\ref{aff127},\ref{aff129}}
\and K.~Ganga\orcid{0000-0001-8159-8208}\inst{\ref{aff1}}
\and J.~Garc\'ia-Bellido\orcid{0000-0002-9370-8360}\inst{\ref{aff115}}
\and T.~Gasparetto\orcid{0000-0002-7913-4866}\inst{\ref{aff7}}
\and V.~Gautard\inst{\ref{aff133}}
\and E.~Gaztanaga\orcid{0000-0001-9632-0815}\inst{\ref{aff63},\ref{aff62},\ref{aff134}}
\and F.~Giacomini\orcid{0000-0002-3129-2814}\inst{\ref{aff23}}
\and F.~Gianotti\orcid{0000-0003-4666-119X}\inst{\ref{aff5}}
\and A.~H.~Gonzalez\orcid{0000-0002-0933-8601}\inst{\ref{aff135}}
\and G.~Gozaliasl\orcid{0000-0002-0236-919X}\inst{\ref{aff136},\ref{aff77}}
\and M.~Guidi\orcid{0000-0001-9408-1101}\inst{\ref{aff22},\ref{aff5}}
\and C.~M.~Gutierrez\orcid{0000-0001-7854-783X}\inst{\ref{aff137}}
\and A.~Hall\orcid{0000-0002-3139-8651}\inst{\ref{aff48}}
\and C.~Hern\'andez-Monteagudo\orcid{0000-0001-5471-9166}\inst{\ref{aff138},\ref{aff47}}
\and H.~Hildebrandt\orcid{0000-0002-9814-3338}\inst{\ref{aff139}}
\and J.~Hjorth\orcid{0000-0002-4571-2306}\inst{\ref{aff93}}
\and J.~J.~E.~Kajava\orcid{0000-0002-3010-8333}\inst{\ref{aff140},\ref{aff141}}
\and Y.~Kang\orcid{0009-0000-8588-7250}\inst{\ref{aff57}}
\and V.~Kansal\orcid{0000-0002-4008-6078}\inst{\ref{aff142},\ref{aff143}}
\and D.~Karagiannis\orcid{0000-0002-4927-0816}\inst{\ref{aff108},\ref{aff144}}
\and K.~Kiiveri\inst{\ref{aff75}}
\and C.~C.~Kirkpatrick\inst{\ref{aff75}}
\and S.~Kruk\orcid{0000-0001-8010-8879}\inst{\ref{aff17}}
\and J.~Le~Graet\orcid{0000-0001-6523-7971}\inst{\ref{aff60}}
\and L.~Legrand\orcid{0000-0003-0610-5252}\inst{\ref{aff145},\ref{aff146}}
\and M.~Lembo\orcid{0000-0002-5271-5070}\inst{\ref{aff108},\ref{aff109}}
\and F.~Lepori\orcid{0009-0000-5061-7138}\inst{\ref{aff147}}
\and G.~Leroy\orcid{0009-0004-2523-4425}\inst{\ref{aff148},\ref{aff87}}
\and G.~F.~Lesci\orcid{0000-0002-4607-2830}\inst{\ref{aff86},\ref{aff5}}
\and J.~Lesgourgues\orcid{0000-0001-7627-353X}\inst{\ref{aff42}}
\and L.~Leuzzi\orcid{0009-0006-4479-7017}\inst{\ref{aff86},\ref{aff5}}
\and T.~I.~Liaudat\orcid{0000-0002-9104-314X}\inst{\ref{aff149}}
\and A.~Loureiro\orcid{0000-0002-4371-0876}\inst{\ref{aff150},\ref{aff151}}
\and J.~Macias-Perez\orcid{0000-0002-5385-2763}\inst{\ref{aff152}}
\and G.~Maggio\orcid{0000-0003-4020-4836}\inst{\ref{aff7}}
\and E.~A.~Magnier\orcid{0000-0002-7965-2815}\inst{\ref{aff45}}
\and C.~Mancini\orcid{0000-0002-4297-0561}\inst{\ref{aff39}}
\and F.~Mannucci\orcid{0000-0002-4803-2381}\inst{\ref{aff153}}
\and R.~Maoli\orcid{0000-0002-6065-3025}\inst{\ref{aff154},\ref{aff43}}
\and C.~J.~A.~P.~Martins\orcid{0000-0002-4886-9261}\inst{\ref{aff155},\ref{aff31}}
\and L.~Maurin\orcid{0000-0002-8406-0857}\inst{\ref{aff12}}
\and M.~Miluzio\inst{\ref{aff17},\ref{aff156}}
\and C.~Moretti\orcid{0000-0003-3314-8936}\inst{\ref{aff21},\ref{aff114},\ref{aff7},\ref{aff8},\ref{aff20}}
\and G.~Morgante\inst{\ref{aff5}}
\and K.~Naidoo\orcid{0000-0002-9182-1802}\inst{\ref{aff134}}
\and A.~Navarro-Alsina\orcid{0000-0002-3173-2592}\inst{\ref{aff84}}
\and S.~Nesseris\orcid{0000-0002-0567-0324}\inst{\ref{aff115}}
\and F.~Passalacqua\orcid{0000-0002-8606-4093}\inst{\ref{aff99},\ref{aff59}}
\and K.~Paterson\orcid{0000-0001-8340-3486}\inst{\ref{aff73}}
\and L.~Patrizii\inst{\ref{aff23}}
\and A.~Pisani\orcid{0000-0002-6146-4437}\inst{\ref{aff60},\ref{aff157}}
\and D.~Potter\orcid{0000-0002-0757-5195}\inst{\ref{aff147}}
\and S.~Quai\orcid{0000-0002-0449-8163}\inst{\ref{aff86},\ref{aff5}}
\and M.~Radovich\orcid{0000-0002-3585-866X}\inst{\ref{aff24}}
\and P.-F.~Rocci\inst{\ref{aff12}}
\and S.~Sacquegna\orcid{0000-0002-8433-6630}\inst{\ref{aff127},\ref{aff128},\ref{aff129}}
\and M.~Sahl\'en\orcid{0000-0003-0973-4804}\inst{\ref{aff158}}
\and D.~B.~Sanders\orcid{0000-0002-1233-9998}\inst{\ref{aff45}}
\and E.~Sarpa\orcid{0000-0002-1256-655X}\inst{\ref{aff21},\ref{aff114},\ref{aff20}}
\and C.~Scarlata\orcid{0000-0002-9136-8876}\inst{\ref{aff159}}
\and A.~Schneider\orcid{0000-0001-7055-8104}\inst{\ref{aff147}}
\and D.~Sciotti\orcid{0009-0008-4519-2620}\inst{\ref{aff43},\ref{aff85}}
\and E.~Sellentin\inst{\ref{aff160},\ref{aff38}}
\and F.~Shankar\orcid{0000-0001-8973-5051}\inst{\ref{aff161}}
\and L.~C.~Smith\orcid{0000-0002-3259-2771}\inst{\ref{aff162}}
\and K.~Tanidis\orcid{0000-0001-9843-5130}\inst{\ref{aff111}}
\and G.~Testera\inst{\ref{aff28}}
\and R.~Teyssier\orcid{0000-0001-7689-0933}\inst{\ref{aff157}}
\and S.~Tosi\orcid{0000-0002-7275-9193}\inst{\ref{aff27},\ref{aff28},\ref{aff10}}
\and A.~Troja\orcid{0000-0003-0239-4595}\inst{\ref{aff99},\ref{aff59}}
\and M.~Tucci\inst{\ref{aff57}}
\and C.~Valieri\inst{\ref{aff23}}
\and A.~Venhola\orcid{0000-0001-6071-4564}\inst{\ref{aff163}}
\and D.~Vergani\orcid{0000-0003-0898-2216}\inst{\ref{aff5}}
\and G.~Verza\orcid{0000-0002-1886-8348}\inst{\ref{aff164}}
\and P.~Vielzeuf\orcid{0000-0003-2035-9339}\inst{\ref{aff60}}
\and N.~A.~Walton\orcid{0000-0003-3983-8778}\inst{\ref{aff162}}}
										   
\institute{Universit\'e Paris Cit\'e, CNRS, Astroparticule et Cosmologie, 75013 Paris, France\label{aff1}
\and
CNRS-UCB International Research Laboratory, Centre Pierre Binetruy, IRL2007, CPB-IN2P3, Berkeley, USA\label{aff2}
\and
Infrared Processing and Analysis Center, California Institute of Technology, Pasadena, CA 91125, USA\label{aff3}
\and
University of California, Los Angeles, CA 90095-1562, USA\label{aff4}
\and
INAF-Osservatorio di Astrofisica e Scienza dello Spazio di Bologna, Via Piero Gobetti 93/3, 40129 Bologna, Italy\label{aff5}
\and
Instituto de Astrof\'isica de Canarias (IAC); Departamento de Astrof\'isica, Universidad de La Laguna (ULL), 38200, La Laguna, Tenerife, Spain\label{aff6}
\and
INAF-Osservatorio Astronomico di Trieste, Via G. B. Tiepolo 11, 34143 Trieste, Italy\label{aff7}
\and
IFPU, Institute for Fundamental Physics of the Universe, via Beirut 2, 34151 Trieste, Italy\label{aff8}
\and
Department of Physics and Astronomy, University of British Columbia, Vancouver, BC V6T 1Z1, Canada\label{aff9}
\and
INAF-Osservatorio Astronomico di Brera, Via Brera 28, 20122 Milano, Italy\label{aff10}
\and
Universit\'e C\^{o}te d'Azur, Observatoire de la C\^{o}te d'Azur, CNRS, Laboratoire Lagrange, Bd de l'Observatoire, CS 34229, 06304 Nice cedex 4, France\label{aff11}
\and
Universit\'e Paris-Saclay, CNRS, Institut d'astrophysique spatiale, 91405, Orsay, France\label{aff12}
\and
Department of Physics and Astronomy, University of California, Davis, CA 95616, USA\label{aff13}
\and
Department of Astronomy, University of Massachusetts, Amherst, MA 01003, USA\label{aff14}
\and
Universit\'e de Strasbourg, CNRS, Observatoire astronomique de Strasbourg, UMR 7550, 67000 Strasbourg, France\label{aff15}
\and
INAF-Osservatorio Astronomico di Brera, Via Brera 28, 20122 Milano, Italy, and INFN-Sezione di Genova, Via Dodecaneso 33, 16146, Genova, Italy\label{aff16}
\and
ESAC/ESA, Camino Bajo del Castillo, s/n., Urb. Villafranca del Castillo, 28692 Villanueva de la Ca\~nada, Madrid, Spain\label{aff17}
\and
School of Mathematics and Physics, University of Surrey, Guildford, Surrey, GU2 7XH, UK\label{aff18}
\and
Universit\'e Paris-Saclay, Universit\'e Paris Cit\'e, CEA, CNRS, AIM, 91191, Gif-sur-Yvette, France\label{aff19}
\and
INFN, Sezione di Trieste, Via Valerio 2, 34127 Trieste TS, Italy\label{aff20}
\and
SISSA, International School for Advanced Studies, Via Bonomea 265, 34136 Trieste TS, Italy\label{aff21}
\and
Dipartimento di Fisica e Astronomia, Universit\`a di Bologna, Via Gobetti 93/2, 40129 Bologna, Italy\label{aff22}
\and
INFN-Sezione di Bologna, Viale Berti Pichat 6/2, 40127 Bologna, Italy\label{aff23}
\and
INAF-Osservatorio Astronomico di Padova, Via dell'Osservatorio 5, 35122 Padova, Italy\label{aff24}
\and
Space Science Data Center, Italian Space Agency, via del Politecnico snc, 00133 Roma, Italy\label{aff25}
\and
INAF-Osservatorio Astrofisico di Torino, Via Osservatorio 20, 10025 Pino Torinese (TO), Italy\label{aff26}
\and
Dipartimento di Fisica, Universit\`a di Genova, Via Dodecaneso 33, 16146, Genova, Italy\label{aff27}
\and
INFN-Sezione di Genova, Via Dodecaneso 33, 16146, Genova, Italy\label{aff28}
\and
Department of Physics "E. Pancini", University Federico II, Via Cinthia 6, 80126, Napoli, Italy\label{aff29}
\and
INAF-Osservatorio Astronomico di Capodimonte, Via Moiariello 16, 80131 Napoli, Italy\label{aff30}
\and
Instituto de Astrof\'isica e Ci\^encias do Espa\c{c}o, Universidade do Porto, CAUP, Rua das Estrelas, PT4150-762 Porto, Portugal\label{aff31}
\and
Faculdade de Ci\^encias da Universidade do Porto, Rua do Campo de Alegre, 4150-007 Porto, Portugal\label{aff32}
\and
Aix-Marseille Universit\'e, CNRS, CNES, LAM, Marseille, France\label{aff33}
\and
Dipartimento di Fisica, Universit\`a degli Studi di Torino, Via P. Giuria 1, 10125 Torino, Italy\label{aff34}
\and
INFN-Sezione di Torino, Via P. Giuria 1, 10125 Torino, Italy\label{aff35}
\and
European Space Agency/ESTEC, Keplerlaan 1, 2201 AZ Noordwijk, The Netherlands\label{aff36}
\and
Institute Lorentz, Leiden University, Niels Bohrweg 2, 2333 CA Leiden, The Netherlands\label{aff37}
\and
Leiden Observatory, Leiden University, Einsteinweg 55, 2333 CC Leiden, The Netherlands\label{aff38}
\and
INAF-IASF Milano, Via Alfonso Corti 12, 20133 Milano, Italy\label{aff39}
\and
Centro de Investigaciones Energ\'eticas, Medioambientales y Tecnol\'ogicas (CIEMAT), Avenida Complutense 40, 28040 Madrid, Spain\label{aff40}
\and
Port d'Informaci\'{o} Cient\'{i}fica, Campus UAB, C. Albareda s/n, 08193 Bellaterra (Barcelona), Spain\label{aff41}
\and
Institute for Theoretical Particle Physics and Cosmology (TTK), RWTH Aachen University, 52056 Aachen, Germany\label{aff42}
\and
INAF-Osservatorio Astronomico di Roma, Via Frascati 33, 00078 Monteporzio Catone, Italy\label{aff43}
\and
INFN section of Naples, Via Cinthia 6, 80126, Napoli, Italy\label{aff44}
\and
Institute for Astronomy, University of Hawaii, 2680 Woodlawn Drive, Honolulu, HI 96822, USA\label{aff45}
\and
Dipartimento di Fisica e Astronomia "Augusto Righi" - Alma Mater Studiorum Universit\`a di Bologna, Viale Berti Pichat 6/2, 40127 Bologna, Italy\label{aff46}
\and
Instituto de Astrof\'{\i}sica de Canarias, V\'{\i}a L\'actea, 38205 La Laguna, Tenerife, Spain\label{aff47}
\and
Institute for Astronomy, University of Edinburgh, Royal Observatory, Blackford Hill, Edinburgh EH9 3HJ, UK\label{aff48}
\and
Jodrell Bank Centre for Astrophysics, Department of Physics and Astronomy, University of Manchester, Oxford Road, Manchester M13 9PL, UK\label{aff49}
\and
European Space Agency/ESRIN, Largo Galileo Galilei 1, 00044 Frascati, Roma, Italy\label{aff50}
\and
Universit\'e Claude Bernard Lyon 1, CNRS/IN2P3, IP2I Lyon, UMR 5822, Villeurbanne, F-69100, France\label{aff51}
\and
Institut de Ci\`{e}ncies del Cosmos (ICCUB), Universitat de Barcelona (IEEC-UB), Mart\'{i} i Franqu\`{e}s 1, 08028 Barcelona, Spain\label{aff52}
\and
Instituci\'o Catalana de Recerca i Estudis Avan\c{c}ats (ICREA), Passeig de Llu\'{\i}s Companys 23, 08010 Barcelona, Spain\label{aff53}
\and
UCB Lyon 1, CNRS/IN2P3, IUF, IP2I Lyon, 4 rue Enrico Fermi, 69622 Villeurbanne, France\label{aff54}
\and
Departamento de F\'isica, Faculdade de Ci\^encias, Universidade de Lisboa, Edif\'icio C8, Campo Grande, PT1749-016 Lisboa, Portugal\label{aff55}
\and
Instituto de Astrof\'isica e Ci\^encias do Espa\c{c}o, Faculdade de Ci\^encias, Universidade de Lisboa, Campo Grande, 1749-016 Lisboa, Portugal\label{aff56}
\and
Department of Astronomy, University of Geneva, ch. d'Ecogia 16, 1290 Versoix, Switzerland\label{aff57}
\and
INAF-Istituto di Astrofisica e Planetologia Spaziali, via del Fosso del Cavaliere, 100, 00100 Roma, Italy\label{aff58}
\and
INFN-Padova, Via Marzolo 8, 35131 Padova, Italy\label{aff59}
\and
Aix-Marseille Universit\'e, CNRS/IN2P3, CPPM, Marseille, France\label{aff60}
\and
INFN-Bologna, Via Irnerio 46, 40126 Bologna, Italy\label{aff61}
\and
Institut d'Estudis Espacials de Catalunya (IEEC),  Edifici RDIT, Campus UPC, 08860 Castelldefels, Barcelona, Spain\label{aff62}
\and
Institute of Space Sciences (ICE, CSIC), Campus UAB, Carrer de Can Magrans, s/n, 08193 Barcelona, Spain\label{aff63}
\and
School of Physics, HH Wills Physics Laboratory, University of Bristol, Tyndall Avenue, Bristol, BS8 1TL, UK\label{aff64}
\and
Universit\"ats-Sternwarte M\"unchen, Fakult\"at f\"ur Physik, Ludwig-Maximilians-Universit\"at M\"unchen, Scheinerstrasse 1, 81679 M\"unchen, Germany\label{aff65}
\and
Max Planck Institute for Extraterrestrial Physics, Giessenbachstr. 1, 85748 Garching, Germany\label{aff66}
\and
Institute of Theoretical Astrophysics, University of Oslo, P.O. Box 1029 Blindern, 0315 Oslo, Norway\label{aff67}
\and
Jet Propulsion Laboratory, California Institute of Technology, 4800 Oak Grove Drive, Pasadena, CA, 91109, USA\label{aff68}
\and
Felix Hormuth Engineering, Goethestr. 17, 69181 Leimen, Germany\label{aff69}
\and
Technical University of Denmark, Elektrovej 327, 2800 Kgs. Lyngby, Denmark\label{aff70}
\and
Cosmic Dawn Center (DAWN), Denmark\label{aff71}
\and
Institut d'Astrophysique de Paris, UMR 7095, CNRS, and Sorbonne Universit\'e, 98 bis boulevard Arago, 75014 Paris, France\label{aff72}
\and
Max-Planck-Institut f\"ur Astronomie, K\"onigstuhl 17, 69117 Heidelberg, Germany\label{aff73}
\and
NASA Goddard Space Flight Center, Greenbelt, MD 20771, USA\label{aff74}
\and
Department of Physics and Helsinki Institute of Physics, Gustaf H\"allstr\"omin katu 2, 00014 University of Helsinki, Finland\label{aff75}
\and
Universit\'e de Gen\`eve, D\'epartement de Physique Th\'eorique and Centre for Astroparticle Physics, 24 quai Ernest-Ansermet, CH-1211 Gen\`eve 4, Switzerland\label{aff76}
\and
Department of Physics, P.O. Box 64, 00014 University of Helsinki, Finland\label{aff77}
\and
Helsinki Institute of Physics, Gustaf H{\"a}llstr{\"o}min katu 2, University of Helsinki, Helsinki, Finland\label{aff78}
\and
Centre de Calcul de l'IN2P3/CNRS, 21 avenue Pierre de Coubertin 69627 Villeurbanne Cedex, France\label{aff79}
\and
Laboratoire d'etude de l'Univers et des phenomenes eXtremes, Observatoire de Paris, Universit\'e PSL, Sorbonne Universit\'e, CNRS, 92190 Meudon, France\label{aff80}
\and
SKA Observatory, Jodrell Bank, Lower Withington, Macclesfield, Cheshire SK11 9FT, UK\label{aff81}
\and
Dipartimento di Fisica "Aldo Pontremoli", Universit\`a degli Studi di Milano, Via Celoria 16, 20133 Milano, Italy\label{aff82}
\and
INFN-Sezione di Milano, Via Celoria 16, 20133 Milano, Italy\label{aff83}
\and
Universit\"at Bonn, Argelander-Institut f\"ur Astronomie, Auf dem H\"ugel 71, 53121 Bonn, Germany\label{aff84}
\and
INFN-Sezione di Roma, Piazzale Aldo Moro, 2 - c/o Dipartimento di Fisica, Edificio G. Marconi, 00185 Roma, Italy\label{aff85}
\and
Dipartimento di Fisica e Astronomia "Augusto Righi" - Alma Mater Studiorum Universit\`a di Bologna, via Piero Gobetti 93/2, 40129 Bologna, Italy\label{aff86}
\and
Department of Physics, Institute for Computational Cosmology, Durham University, South Road, Durham, DH1 3LE, UK\label{aff87}
\and
Institut d'Astrophysique de Paris, 98bis Boulevard Arago, 75014, Paris, France\label{aff88}
\and
Institute of Physics, Laboratory of Astrophysics, Ecole Polytechnique F\'ed\'erale de Lausanne (EPFL), Observatoire de Sauverny, 1290 Versoix, Switzerland\label{aff89}
\and
Aurora Technology for European Space Agency (ESA), Camino bajo del Castillo, s/n, Urbanizacion Villafranca del Castillo, Villanueva de la Ca\~nada, 28692 Madrid, Spain\label{aff90}
\and
Institut de F\'{i}sica d'Altes Energies (IFAE), The Barcelona Institute of Science and Technology, Campus UAB, 08193 Bellaterra (Barcelona), Spain\label{aff91}
\and
School of Mathematics, Statistics and Physics, Newcastle University, Herschel Building, Newcastle-upon-Tyne, NE1 7RU, UK\label{aff92}
\and
DARK, Niels Bohr Institute, University of Copenhagen, Jagtvej 155, 2200 Copenhagen, Denmark\label{aff93}
\and
Waterloo Centre for Astrophysics, University of Waterloo, Waterloo, Ontario N2L 3G1, Canada\label{aff94}
\and
Department of Physics and Astronomy, University of Waterloo, Waterloo, Ontario N2L 3G1, Canada\label{aff95}
\and
Perimeter Institute for Theoretical Physics, Waterloo, Ontario N2L 2Y5, Canada\label{aff96}
\and
Centre National d'Etudes Spatiales -- Centre spatial de Toulouse, 18 avenue Edouard Belin, 31401 Toulouse Cedex 9, France\label{aff97}
\and
Institute of Space Science, Str. Atomistilor, nr. 409 M\u{a}gurele, Ilfov, 077125, Romania\label{aff98}
\and
Dipartimento di Fisica e Astronomia "G. Galilei", Universit\`a di Padova, Via Marzolo 8, 35131 Padova, Italy\label{aff99}
\and
Institut f\"ur Theoretische Physik, University of Heidelberg, Philosophenweg 16, 69120 Heidelberg, Germany\label{aff100}
\and
Institut de Recherche en Astrophysique et Plan\'etologie (IRAP), Universit\'e de Toulouse, CNRS, UPS, CNES, 14 Av. Edouard Belin, 31400 Toulouse, France\label{aff101}
\and
Universit\'e St Joseph; Faculty of Sciences, Beirut, Lebanon\label{aff102}
\and
Departamento de F\'isica, FCFM, Universidad de Chile, Blanco Encalada 2008, Santiago, Chile\label{aff103}
\and
Satlantis, University Science Park, Sede Bld 48940, Leioa-Bilbao, Spain\label{aff104}
\and
Instituto de Astrof\'isica e Ci\^encias do Espa\c{c}o, Faculdade de Ci\^encias, Universidade de Lisboa, Tapada da Ajuda, 1349-018 Lisboa, Portugal\label{aff105}
\and
Universidad Polit\'ecnica de Cartagena, Departamento de Electr\'onica y Tecnolog\'ia de Computadoras,  Plaza del Hospital 1, 30202 Cartagena, Spain\label{aff106}
\and
Kapteyn Astronomical Institute, University of Groningen, PO Box 800, 9700 AV Groningen, The Netherlands\label{aff107}
\and
Dipartimento di Fisica e Scienze della Terra, Universit\`a degli Studi di Ferrara, Via Giuseppe Saragat 1, 44122 Ferrara, Italy\label{aff108}
\and
Istituto Nazionale di Fisica Nucleare, Sezione di Ferrara, Via Giuseppe Saragat 1, 44122 Ferrara, Italy\label{aff109}
\and
INAF, Istituto di Radioastronomia, Via Piero Gobetti 101, 40129 Bologna, Italy\label{aff110}
\and
Department of Physics, Oxford University, Keble Road, Oxford OX1 3RH, UK\label{aff111}
\and
INAF - Osservatorio Astronomico di Brera, via Emilio Bianchi 46, 23807 Merate, Italy\label{aff112}
\and
ICL, Junia, Universit\'e Catholique de Lille, LITL, 59000 Lille, France\label{aff113}
\and
ICSC - Centro Nazionale di Ricerca in High Performance Computing, Big Data e Quantum Computing, Via Magnanelli 2, Bologna, Italy\label{aff114}
\and
Instituto de F\'isica Te\'orica UAM-CSIC, Campus de Cantoblanco, 28049 Madrid, Spain\label{aff115}
\and
CERCA/ISO, Department of Physics, Case Western Reserve University, 10900 Euclid Avenue, Cleveland, OH 44106, USA\label{aff116}
\and
Technical University of Munich, TUM School of Natural Sciences, Physics Department, James-Franck-Str.~1, 85748 Garching, Germany\label{aff117}
\and
Max-Planck-Institut f\"ur Astrophysik, Karl-Schwarzschild-Str.~1, 85748 Garching, Germany\label{aff118}
\and
Laboratoire Univers et Th\'eorie, Observatoire de Paris, Universit\'e PSL, Universit\'e Paris Cit\'e, CNRS, 92190 Meudon, France\label{aff119}
\and
Departamento de F{\'\i}sica Fundamental. Universidad de Salamanca. Plaza de la Merced s/n. 37008 Salamanca, Spain\label{aff120}
\and
Center for Data-Driven Discovery, Kavli IPMU (WPI), UTIAS, The University of Tokyo, Kashiwa, Chiba 277-8583, Japan\label{aff121}
\and
Ludwig-Maximilians-University, Schellingstrasse 4, 80799 Munich, Germany\label{aff122}
\and
Max-Planck-Institut f\"ur Physik, Boltzmannstr. 8, 85748 Garching, Germany\label{aff123}
\and
Dipartimento di Fisica - Sezione di Astronomia, Universit\`a di Trieste, Via Tiepolo 11, 34131 Trieste, Italy\label{aff124}
\and
California Institute of Technology, 1200 E California Blvd, Pasadena, CA 91125, USA\label{aff125}
\and
Department of Physics \& Astronomy, University of California Irvine, Irvine CA 92697, USA\label{aff126}
\and
Department of Mathematics and Physics E. De Giorgi, University of Salento, Via per Arnesano, CP-I93, 73100, Lecce, Italy\label{aff127}
\and
INFN, Sezione di Lecce, Via per Arnesano, CP-193, 73100, Lecce, Italy\label{aff128}
\and
INAF-Sezione di Lecce, c/o Dipartimento Matematica e Fisica, Via per Arnesano, 73100, Lecce, Italy\label{aff129}
\and
Departamento F\'isica Aplicada, Universidad Polit\'ecnica de Cartagena, Campus Muralla del Mar, 30202 Cartagena, Murcia, Spain\label{aff130}
\and
Instituto de F\'isica de Cantabria, Edificio Juan Jord\'a, Avenida de los Castros, 39005 Santander, Spain\label{aff131}
\and
Observatorio Nacional, Rua General Jose Cristino, 77-Bairro Imperial de Sao Cristovao, Rio de Janeiro, 20921-400, Brazil\label{aff132}
\and
CEA Saclay, DFR/IRFU, Service d'Astrophysique, Bat. 709, 91191 Gif-sur-Yvette, France\label{aff133}
\and
Institute of Cosmology and Gravitation, University of Portsmouth, Portsmouth PO1 3FX, UK\label{aff134}
\and
Department of Astronomy, University of Florida, Bryant Space Science Center, Gainesville, FL 32611, USA\label{aff135}
\and
Department of Computer Science, Aalto University, PO Box 15400, Espoo, FI-00 076, Finland\label{aff136}
\and
Instituto de Astrof\'\i sica de Canarias, c/ Via Lactea s/n, La Laguna 38200, Spain. Departamento de Astrof\'\i sica de la Universidad de La Laguna, Avda. Francisco Sanchez, La Laguna, 38200, Spain\label{aff137}
\and
Universidad de La Laguna, Departamento de Astrof\'{\i}sica, 38206 La Laguna, Tenerife, Spain\label{aff138}
\and
Ruhr University Bochum, Faculty of Physics and Astronomy, Astronomical Institute (AIRUB), German Centre for Cosmological Lensing (GCCL), 44780 Bochum, Germany\label{aff139}
\and
Department of Physics and Astronomy, Vesilinnantie 5, 20014 University of Turku, Finland\label{aff140}
\and
Serco for European Space Agency (ESA), Camino bajo del Castillo, s/n, Urbanizacion Villafranca del Castillo, Villanueva de la Ca\~nada, 28692 Madrid, Spain\label{aff141}
\and
ARC Centre of Excellence for Dark Matter Particle Physics, Melbourne, Australia\label{aff142}
\and
Centre for Astrophysics \& Supercomputing, Swinburne University of Technology,  Hawthorn, Victoria 3122, Australia\label{aff143}
\and
Department of Physics and Astronomy, University of the Western Cape, Bellville, Cape Town, 7535, South Africa\label{aff144}
\and
DAMTP, Centre for Mathematical Sciences, Wilberforce Road, Cambridge CB3 0WA, UK\label{aff145}
\and
Kavli Institute for Cosmology Cambridge, Madingley Road, Cambridge, CB3 0HA, UK\label{aff146}
\and
Department of Astrophysics, University of Zurich, Winterthurerstrasse 190, 8057 Zurich, Switzerland\label{aff147}
\and
Department of Physics, Centre for Extragalactic Astronomy, Durham University, South Road, Durham, DH1 3LE, UK\label{aff148}
\and
IRFU, CEA, Universit\'e Paris-Saclay 91191 Gif-sur-Yvette Cedex, France\label{aff149}
\and
Oskar Klein Centre for Cosmoparticle Physics, Department of Physics, Stockholm University, Stockholm, SE-106 91, Sweden\label{aff150}
\and
Astrophysics Group, Blackett Laboratory, Imperial College London, London SW7 2AZ, UK\label{aff151}
\and
Univ. Grenoble Alpes, CNRS, Grenoble INP, LPSC-IN2P3, 53, Avenue des Martyrs, 38000, Grenoble, France\label{aff152}
\and
INAF-Osservatorio Astrofisico di Arcetri, Largo E. Fermi 5, 50125, Firenze, Italy\label{aff153}
\and
Dipartimento di Fisica, Sapienza Universit\`a di Roma, Piazzale Aldo Moro 2, 00185 Roma, Italy\label{aff154}
\and
Centro de Astrof\'{\i}sica da Universidade do Porto, Rua das Estrelas, 4150-762 Porto, Portugal\label{aff155}
\and
HE Space for European Space Agency (ESA), Camino bajo del Castillo, s/n, Urbanizacion Villafranca del Castillo, Villanueva de la Ca\~nada, 28692 Madrid, Spain\label{aff156}
\and
Department of Astrophysical Sciences, Peyton Hall, Princeton University, Princeton, NJ 08544, USA\label{aff157}
\and
Theoretical astrophysics, Department of Physics and Astronomy, Uppsala University, Box 515, 751 20 Uppsala, Sweden\label{aff158}
\and
Minnesota Institute for Astrophysics, University of Minnesota, 116 Church St SE, Minneapolis, MN 55455, USA\label{aff159}
\and
Mathematical Institute, University of Leiden, Einsteinweg 55, 2333 CA Leiden, The Netherlands\label{aff160}
\and
School of Physics \& Astronomy, University of Southampton, Highfield Campus, Southampton SO17 1BJ, UK\label{aff161}
\and
Institute of Astronomy, University of Cambridge, Madingley Road, Cambridge CB3 0HA, UK\label{aff162}
\and
Space physics and astronomy research unit, University of Oulu, Pentti Kaiteran katu 1, FI-90014 Oulu, Finland\label{aff163}
\and
Center for Computational Astrophysics, Flatiron Institute, 162 5th Avenue, 10010, New York, NY, USA\label{aff164}}


%
%
 \abstract{The \Euclid spacecraft will detect tens of thousands of clusters and protoclusters at  $z>1.3$ over the course of its mission. With a total coverage of $63.1\,\mathrm{deg}^2$, the Euclid Quick Data Release 1 (Q1) is large enough to detect tens of clusters and hundreds of protoclusters at these early epochs. The Q1 photometric redshift catalogue enables us to detect clusters out to $z \lesssim 1.5$; however, infrared imaging from the \Spitzer extends this limit to higher redshifts by using high local projected densities of \spitzer-selected galaxies as signposts for cluster and protocluster candidates. We use \spitzer imaging of the \acrlong{edf} (\acrshort{edf}) to derive densities for a sample of \spitzer-selected galaxies at redshifts $z > 1.3$, building \spitzer IRAC1 and IRAC2 photometric catalogues that are 95\% complete at a magnitude limit of ${\rm IRAC2}=$22.2, 22.6, and 22.8 for the \acrlong{edfs}, \acrlong{edff}, and \acrlong{edfn}, respectively. 
 We apply two complementary methods to calculate galaxy densities: (1) aperture and surface density; and (2) the $N$th-nearest-neighbour method.  When considering a sample selected at a magnitude limit of ${\rm IRAC2} < 22.2$, at which all three \acrshort{edf} are 95\% complete, our surface density distributions are consistent among the three \acrshort{edf} and with the United Kingdom Infrared Telescope Infrared Deep Sky Survey Ultra-Deep Survey blank field survey. We also considered a deeper sample at a magnitude limit of ${\rm IRAC2} < 22.8$, finding that 2\% and 3\% of the surface densities in the North and Fornax fields are $3\,\sigma$ higher than the average field distribution and similar to densities found in the Clusters Around Active Galactic Nuclei cluster survey. Our surface densities are also consistent with predictions from the \acrlong{gaea} semi-analytical model. Using combined \Euclid and ground-based $i$-band photometry from the \acrlong{dawn}, we show that our highest \spitzer-selected galaxy overdence regions, found at $z \approx 1.5$, also host high densities of passive galaxies. This means that we measure densities consistent with those found in clusters and protoclusters at $z>1.3$, and our catalogues will allow us to extend cluster and protocluster detections to $z>1.3$ in the \acrshort{edf}.
}
%
%
%
%
   \titlerunning{\Euclid\/: Euclid and Spitzer galaxy density catalogues at z$>$1.3 }
   \authorrunning{Euclid Collaboration: N. Mai et al.}
   
  \maketitle
%
%
%
%
   
\section{\label{sc:Intro}Introduction}
Galaxy clusters are ideal for studying the interactions between galaxies and their environment, enabling us to quantify the impact of local environment on the evolution of galaxy properties.
At redshifts out to $z\approx 1$, the relation between star-formation rate and environment indicates that massive galaxies in dense regions, such as those found in galaxy clusters, tend to suppress their star formation, while more isolated galaxies exhibit higher star-formation rates \citep[e.g.,][]{Gomez03, Mei09, Peng2010, Peng2012, Lemaux19}. At present, it is still unclear how this relation behaves at higher redshifts. In fact, different studies have identified cluster cores dominated by star-forming galaxies \citep[e.g.,][]{Tran10, Fassbender11,Hayashi11, Tadaki12,Zeimann2012,Brodwin13,Mei15,Alberts16,Hayashi16,Shimakawa18,Aoyama22,Koyama21,Polletta21,Zheng21} but also cores dominated by passive galaxies \citep[e.g.,][]{Andreon14, Cooke15, Strazzullo13, Noirot16, Noirot18,Markov20,Sazonova20, Mei23}.  Some clusters show both populations \citep{Wang16,Kubo17,Strazzullo18}, and others present starbursts \citep{Casey15,Casey16,Wang16}. 

Similarly, galaxy morphological type is strongly correlated with galaxy environment. Observations at low redshifts ($z < 0.5$) show that clusters predominantly host massive early-type galaxies that have evolved passively since redshifts of $z \approx 2-3$ \citep{Stanford98,vanDokkum07, Mei09},
while late-type galaxies are more prominent in isolated regions. This suggests a correlation between galaxy morphology and the surrounding environment, the so-called morphology-density relation (e.g., \citealt{Postman05, Dressler80, Mei23}). \citet{Mei23} investigated and confirmed this relation out to $z\approx 2$; however, its behaviour at higher redshifts remains uncertain due to the paucity of statistical cluster samples at these redshifts.

The \Euclid\ mission \citep{EuclidSkyOverview} will dramatically advance studies of clusters and protoclusters (the groups that eventually merge into galaxy clusters by the present) by detecting tens of thousands of clusters (with masses $M>10^{14}M_\odot$) and protoclusters at $z>1.3$ \citep[e.g., ][]{Sartoris2016,Ascaso17}, a still poorly explored redshift range when the first structures in the Universe form. These detections will benefit from \Euclid's high-resolution infrared imaging and grism spectroscopy, as well as multiwavelength ancillary imaging and spectroscopy. The \gls{q1} covers a combined area of 63.1~$\mathrm{deg}^2$ \citep{Q1-TP001}, which is large enough to detect tens of clusters and hundreds of protoclusters at these redshifts \citep{Sartoris2016}. Large uncertainties in the \gls{q1} photometric redshifts restrict cluster detection to $z \leq 1.5$ (Bhargava et al., in prep.). However, combined space-based infrared imaging from \Euclid and the \Spitzer extends this limit to higher redshift \citep{Wylezalek13, Wylezalek14, Noirot18, Mei23}.


In fact, the \acrlong{irac} \citep[\acrshort{irac},][]{Fazio04} has played a key role in detecting galaxy clusters across a wide range of masses and redshifts. In particular, \spitzer-selected galaxy overdense regions have been successfully used to identify galaxy clusters and protoclusters at redshift $z>1.3$ \citep{Papovich08,Wylezalek13, Wylezalek14,Rettura14,Baronchelli16,Greenslade18,Martinache18,Noirot16, Noirot18, Mei23,Gully24}, using a colour selection in IRAC channel~1 ($\lambda =3.6\,\micron$; hereafter IRAC1) and channel~2 ($\lambda = 4.5\,\micron$; hereafter IRAC2). 
This colour cut combined with an \acrshort{irac} magnitude limit selects samples of massive galaxies at redshift $z>1.3$ that are approxinatively 95\% complete and 95\% pure \citep{Mei23} up to a given mass limit, regardless of their morphological type or star-formation activity.  

In this paper, we complement the \gls{q1} data with public archival \spitzer data. Our goal is to identify  \spitzer-selected galaxy overdense regions at $z > 1.3$ and prepare for the detection of cluster and protocluster candidates with \Euclid at these redshifts. We measure \spitzer photometry in the \gls{q1} fields, calculate local projected galaxy densities, and publish examples of the highest-density regions that we find. Using \Euclid and ground-based $i$-band photometry, we select passive galaxies in the same fields, and we give examples of some of our highest-density regions at $z\sim1.5$ that also host high densities of passive galaxies, characteristic of galaxy clusters. This demonstrates the potential of our density catalogs for future detections of clusters and protoclusters at $z>1.3$ in the \gls{q1} fields.  

The structure of the paper is as follows. \Cref{sc:Observations} provides an overview of the \spitzer, \Euclid, and ground-based observations, and of the \gls{gaea} semi-analytical model. \Cref{sc:Photometry} describes our \spitzer photometric measurements. \Cref{sc:Density} presents the two methods used to compute the local projected galaxy densities, the final catalogues, and our density distributions.  \Cref{sec:results} presents our results and compares them to \spitzer-selected galaxy density measurements in a blank field and in a cluster survey, as well as to the predictions from the \gls{gaea} model \citep{Delucia07, Delucia24}. Finally, \cref{sec:discussion} compares our density measurements to other density detections in the EDF. 

Unless otherwise specified, we adopt the \Planck~2015 \citep{Planck15} flat $\Lambda$CDM cosmology, with $\Omega_{\mathrm{m}}=0.308$,
 $\Omega_{\Lambda} =0.692$, and $H_0=67.8\,\kmsMpc$; magnitudes are given in the AB system \citep{Oke83,Sirianni05}.

\section{\label{sc:Observations} Observations and simulations}

\subsection{The \Euclid Q1 release and ancillary data}
The \gls{q1} release \citep{Q1-TP001} focuses on the three \gls{edf}, namely, the \gls{edff}, \gls{edfn}, and \gls{edfs}.
These fields were selected to minimise contamination from foreground sources, allowing for the study of high-redshift objects. The \gls{edff} covers 12.1\,$\mathrm{deg}^2$, the \gls{edfn}, 22.9\,$\mathrm{deg}^2$, and the \gls{edfs}, 28.1\,$\mathrm{deg}^2$. 

The Q1 data set consists of 351 tiles, each with a field of view of $0.57\,\mathrm{deg}^2$, corresponding to $ \ang{0.75;;}\times \ang{0.75;;}$. These observations reach a $10~\sigma$ magnitude limit of 24.5 in the \gls{vis} filter, \IE \citep{EuclidSkyVIS}, and a $5~\sigma$ magnitude limit of 24.5 in each of the filters of the \gls{nisp}: \YE, \JE, and \HE \citep{EuclidSkyNISP}. The spatial resolution of the images is defined by a pixel scale of \ang{;;0.1} for \gls{vis} and \ang{;;0.3} for \gls{nisp}. We focus on \Euclid infrared observations in the \HE filter. 

The \gls{q1} release includes galaxy photometry from \Euclid and external data, and galaxy properties obtained with the \Euclid \gls{sgs} \gls{ou} MER pipeline \citep{Q1-TP004}. We use the \gls{q1} catalogue photometric redshifts derived by the \texttt{Phosphorus} method, which were produced by OU-PHZ \citep{Q1-TP005}.  This method provides Bayesian posterior distributions of photometric redshifts. 
To ensure consistent data quality, we selected galaxies from the \gls{q1} catalogue by applying detection quality flags generated by the pipeline to filter out sources that are saturated, located near image borders, affected by contamination from nearby objects, blended by other sources, or classified as spurious detections. 

For the ancillary data, we use ground-based observations of the \gls{edfn} and \gls{edff} from the \gls{hsc} \textit{i}-band (hereafter $\ihsc$) from the \gls{h20}, as part of the \acrlong{dawn} \citep[\acrshort{dawn};][]{Zalesky24}. The \acrshort{dawn} photometric catalogue was obtained using \texttt{The Farmer} \citep{Weaver23}, a software package designed to recover fluxes by modelling surface brightness profiles. The $5~\sigma$ magnitude depth is $\ihsc=25.9$. 


\subsection{\Spitzer observations}
We use public \spitzer IRAC1 and IRAC2 mosaic images from \citet{Moneti22}. These images were created by combining archival data from both the pre- and post-cryogenic missions, along with legacy surveys specifically designed to enhance the coverage in the \gls{edf}. The program IDs and details are found in \citet{Moneti22}. 

The IRAC1 and IRAC2 images each cover a $\ang{;5.2;} \times \ang{;5.2;}$ field of view.
The \gls{irac} \gls{psf} has a full width at half maximum of \ang{;;1.95} and \ang{;;2.02} in IRAC1 and IRAC2, respectively (IRAC Instrument Handbook\footnote{\url{https://irsa.ipac.caltech.edu/data/SPITZER/docs/irac/iracinstrumenthandbook/5/}}). 
The observations were processed and mosaiced using the \texttt{MOPEX} package \citep{Makovoz05} and resampled to a pixel scale of \ang{;;0.6}. 

The \gls{edfs} is covered by uniform observations obtained 
over an area of $\approx23.37\,\mathrm{deg}^2$ in both channels. The  \gls{edfn} and \gls{edff} include multiple reprocessed archival data sets with varying observation strategies and characteristics, resulting in inhomogeneous depth and coverage. They cover an area of approximately $11.64\,\mathrm{deg}^2$ and $10.79\,\mathrm{deg}^2$, respectively. 
The depth of the IRAC1 images reaches at least $24\,\mathrm{mag}$ at $5~\sigma$ in a \ang{;;2.5} aperture. Details of the image processing can be found in \citet{Moneti22}. 

\subsection{Simulations}
\label{sec:gaea}
The \gls{gaea} simulations are based on a semi-analytical model \citep{Delucia07, Delucia24}, designed to study galaxy formation and evolution in a cosmological context by including explicit modelling of the relevant physical processes governing the evolution of the baryonic components. It is built on merger trees derived from the Millenium Simulation \citep{Springel05}, which follow the hierarchical growth of dark matter halos and provide the structural framework for modelling baryonic processes, such as \gls{agn} feedback, disc instabilities, reheating and ejection efficiency in stellar feedback, and gas ram-pressure stripping from satellite galaxies. 

In the following, we use a simulated light-cone that has been created for the Euclid Consortium and is based on the latest version of the \gls{gaea} model \citep{Delucia24}. The light cone covers a \ang{5.27;;} diameter aperture, and includes approximately $\num{6500000}$ galaxies within the redshift range $0 \leq z \leq 4$. In addition to a number of predicted physical properties, the light cone includes photometry in a large set of photometric bands, including IRAC1 and IRAC2. Although the cosmology used in \gls{gaea} differs from that used in this paper, we do not expect a significant impact on the results when comparing this model to observations (see \cref{sec:galover}). 

\section{\label{sc:Photometry} Photometry}
\subsection{ \label{sc:mask} Bright source masking }

Source detection was performed using the \sextractor software \citep{Bertin96}, a widely-used tool for photometric analysis. 
We mask bright sources before performing photometric measurements on the \spitzer images. As noted by \citet{Ji18}, accurate extraction of faint sources requires precise background estimation. Given that our fields contain some bright sources, it is crucial to mask these objects to allow \sextractor (see \cref{sc:sextractor}) to compute the background more accurately. To achieve this, we perform an initial extraction in a so-called `cold mode', where we identify only the brightest sources using a high detection threshold. The output aperture maps are then used to mask sources with magnitudes brighter than 17. 

To account for extended or elongated objects, we employ the Kron aperture to define the mask. Although the aperture computed by \sextractor is generally sufficient, the brightest sources tend to have artifacts or to be highly extended, which contaminates the background estimation. Considering this, we follow the approach outlined by \citet{Kelvin23} and expand the mask size by adding a number of pixels $p$ to our mask radii, with
\begin{equation}
p \equiv 10^{-0.2 \, (m_{\rm IRAC} - 17)} + 6 \, ,
\end{equation}
where $m_{\rm IRAC}$ is our catalogue magnitude computed by \sextractor. 
These additional pixels increase the mask size as the magnitude decreases, encompassing the extended flux of bright sources that could otherwise perturb the local background estimation and thus contaminate the flux of nearby faint sources. As a result of these masking procedures, the total area of each field is slightly reduced. We report an average areal reduction of 2.6\% for the three regions through the two channels.

\subsection{Source extraction}

\subsubsection{\sextractor} \label{sc:sextractor}
To ensure consistent flux determination across both IRAC channels, we use \sextractor in dual mode, with IRAC1 as the detection image, and measure \verb|MAG_AUTO| magnitudes.
The images are in units of MJy$\,\mathrm{sr}^{-1}$, hence we apply a zero-point magnitude of 21.58 to convert them to AB magnitudes\footnote{Considering the \ang{;;0.6} pixel size of \textit{Spitzer}'s images.}. Weight maps are provided in the  public data release.  We give our \sextractor configuration parameters in \cref{tab:sexpar}, optimized by  \citet{Lacy05} for faint source detection, such as needed for high-redshift galaxies. 
The final photometric uncertainties are the sum in quadrature of the statistical uncertainty, the shot noise, and the uncertainty on the photometric zero point.  


\begin{figure}[t!]
\centering
\includegraphics[angle=0,width=1\hsize]{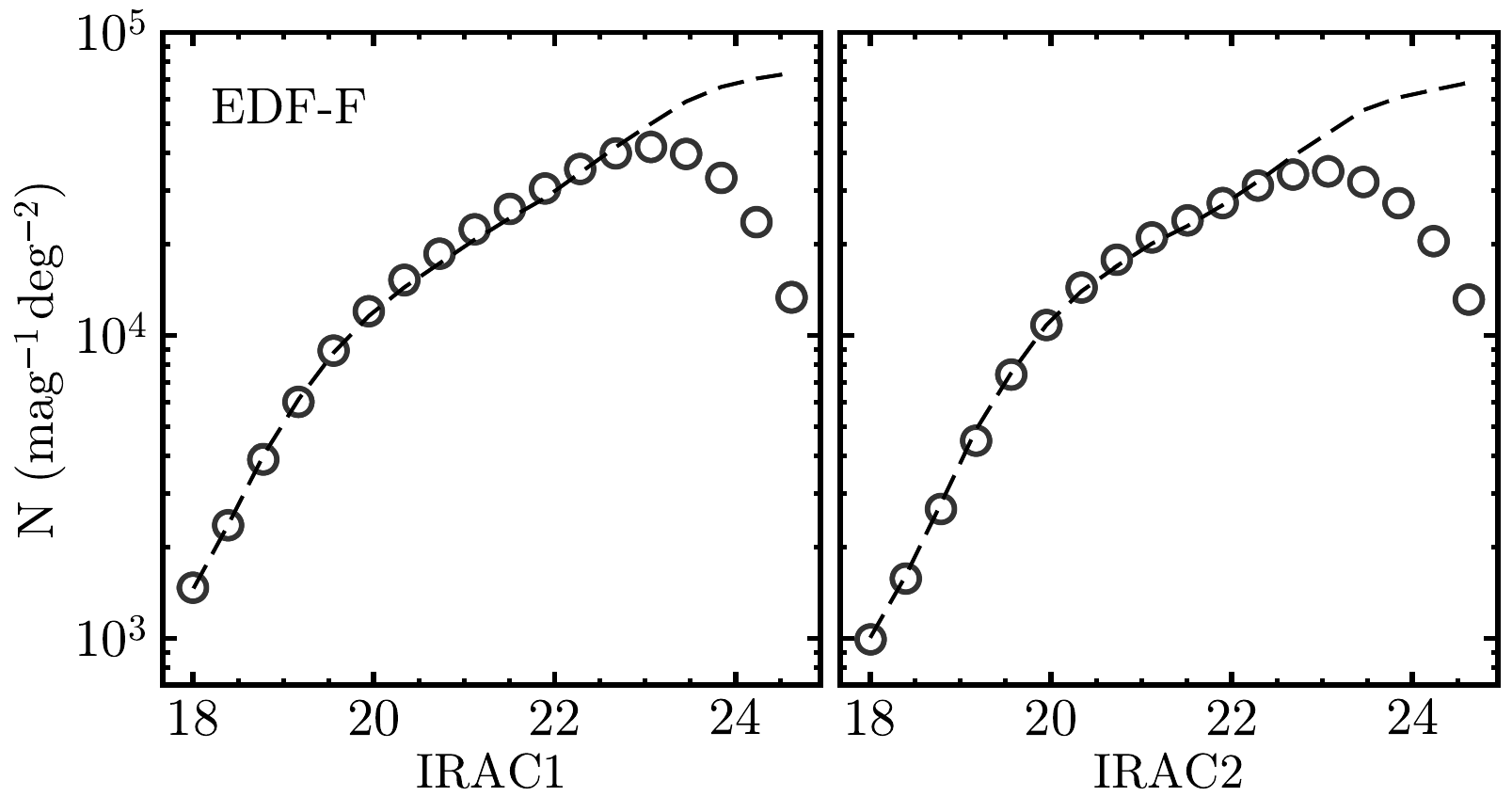}
\includegraphics[angle=0,width=1\hsize]{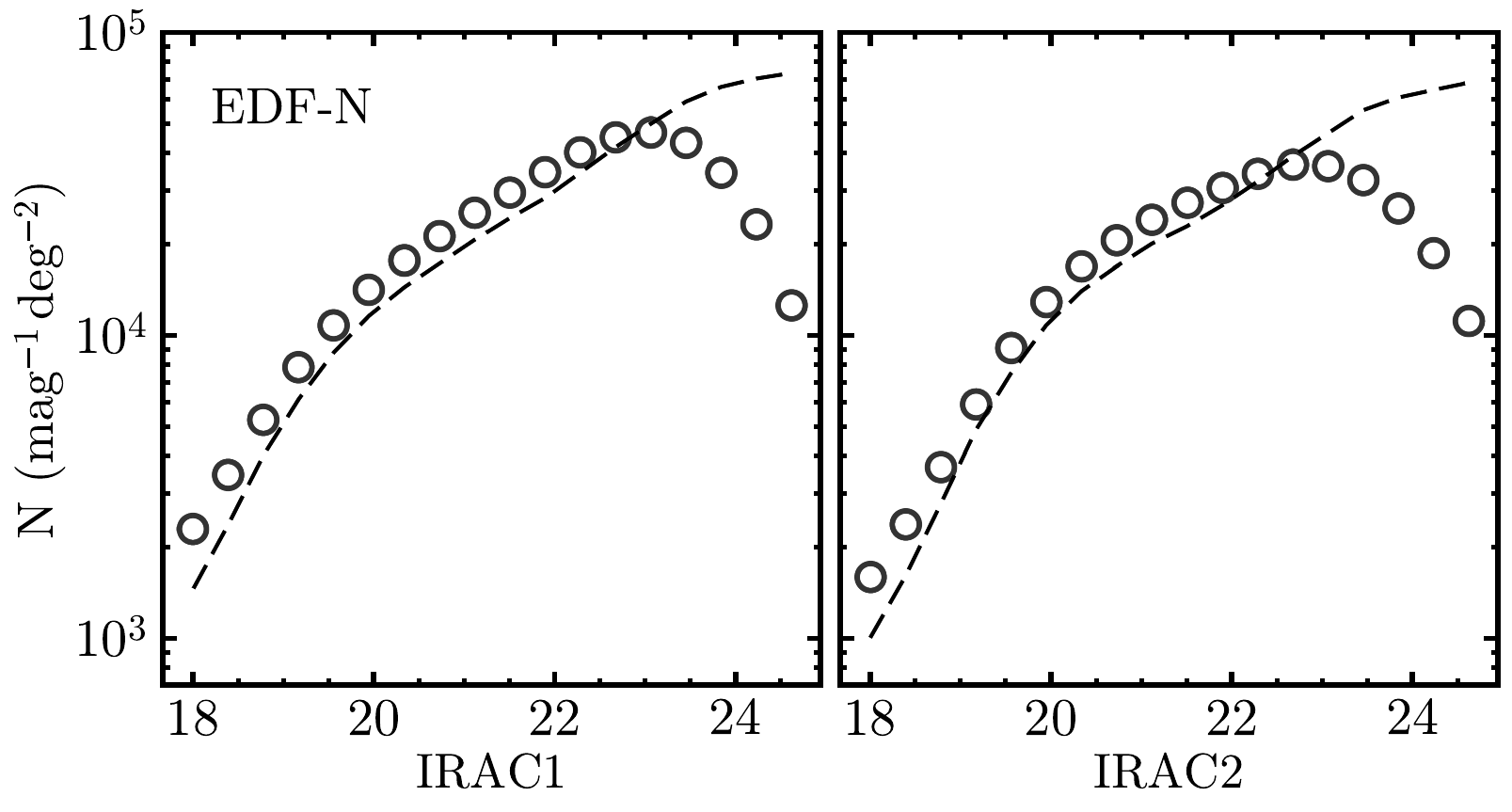}
\includegraphics[angle=0,width=1\hsize]{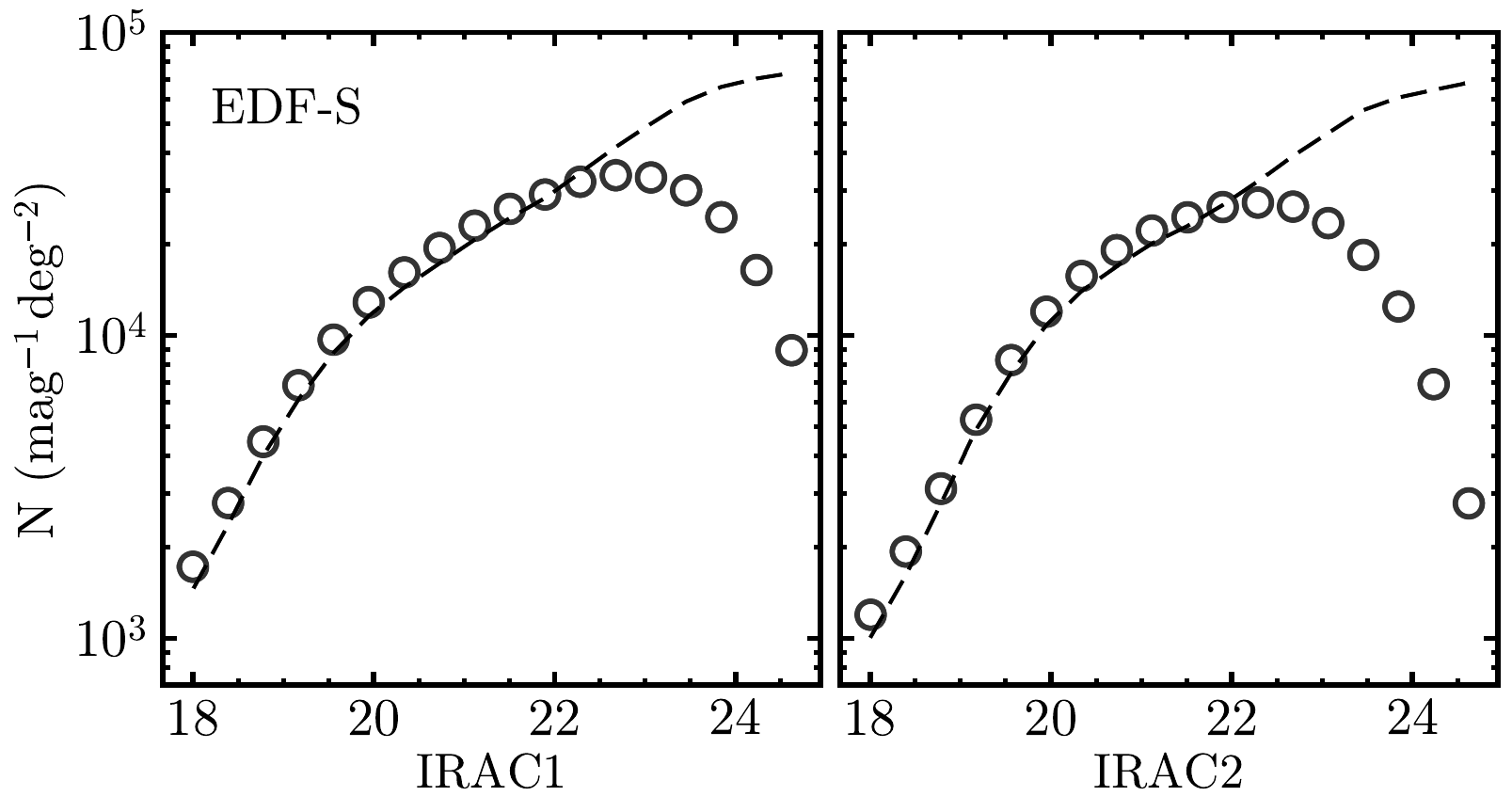}
\caption{Our catalogue number counts. We show the number of detections as a function of magnitude. The circles represent our data, while the black dashed line corresponds to the COSMOS2020 catalogue. Our catalogues yield similar number counts for \gls{edff} and \gls{edfn}. However, \gls{edfs} displays a slight deficit of faint sources, primarily because of the shallower magnitude limit. Our results are consistent with those of \citet{Moneti22}.}
\label{fig:1}
\end{figure}

\Cref{fig:1} illustrates the number of detected sources as a function of IRAC1 and IRAC2 magnitudes. To asses the completeness of our catalogues, we compare our number counts to that from COSMOS2020 \citep{Weaver22}, which has deeper \spitzer observations. We compute our catalogue completeness in magnitude bins by dividing our number counts by the COSMOS2020 one. Table~\ref{tab:completeness} gives the magnitude limits at which our photometric catalogues are 95\% complete for each \gls{q1} field. 

Our results qualitatively agree with the \citet{Moneti22} number counts shown in their figure~8. However, we could not perform a quantitative analysis because their catalogues are not public. Also, while comparing our catalogues to COSMOS2020, we assume that COSMOS2020 is a representative catalogue for galaxy number counts, and do not take into account field-to-field variations due to sampling variance. 

\begin{table}[!t]
    \caption{Magnitude limits corresponding to 95\% completeness for the three \gls{edf}.}
    \label{tab:completeness}
    \centering
    \setlength{\tabcolsep}{8pt} 
    \begin{tabular}{lcc}
        \hline\hline\T\B
        Field & IRAC1 & IRAC2 \\
        \hline
        \gls{edfs}  & 22.4  & 22.2 \T \\
        \gls{edff}  & 22.9  & 22.6  \T \\
        \gls{edfn}  & 23.2  & 22.8  \T \B  \\
        \hline
    \end{tabular}
\end{table}

\subsubsection{TPHOT \spitzer photometry}

To improve deblending of \spitzer galaxies, and therefore attain more precise photometry, we perform additional photometric measurements using the \tphot software \citep{Merlin15, Merlin16}. This is specifically designed to perform photometry while accounting for source blending using higher resolution images as a reference to obtain photometry in lower resolution images of the same field. It incorporates spatial and morphological information extracted from the high-resolution image to accurately model and measure the fluxes of sources in a lower-resolution image. 

The core of the \tphot methodology relies on constructing and solving a linear system that minimises the $\chi^2$ comparison between the observed low-resolution image pixel values and the model fluxes derived from the high-resolution priors. In this paper, we use high-resolution \Euclid \HE images as priors to measure photometry in IRAC images. We give our \tphot parameters in Table~\ref{tab:tphotpar}. 

The \Euclid input is the \gls{q1} photometric catalogue and its associated segmentation maps \citep{Q1-TP004}. We then apply the \tphot pipeline using the cell-on-object fitting method to obtain the multiwavelength photometric catalogue. The statistical uncertainty on the photometry was estimated with \tphot. \citet{Mei23} have shown that for galaxies at our redshifts, Monte-Carlo simulations confirm \tphot estimates. The final photometric uncertainties are the sum in quadrature of the statistical uncertainty, shot noise, and the uncertainty on the photometric zero point. 



\begin{table}[!t]
    \caption{Average background aperture density $N_{\rm bkg}$ calculated at a radius $R$, and its standard deviation $\sigma_{\rm bkg}$, calculated in the \acrshort{spuds} field, at the depth of S1 and S2.}
    \label{tab:bkg}
    \centering
    \setlength{\tabcolsep}{8pt} 
    \begin{tabular}{lcccc}
        \hline\hline\T\B
        Field &Sample& $R$&$N_{\rm bkg} \pm \sigma_{\rm bkg}$ \\
        \hline
        \noalign{\vskip 2pt}
      SpUDS& S1  & \ang{;0.5;}&1.7 $\pm$ 0.6  \\
       && \ang{;1;}& 7.0 $\pm$ 2.2  \\
        &S2 &\ang{;0.5;} & 2.4  $\pm$ 0.7   \\
        && \ang{;1;}& 9.6 $\pm$ 2.1  \\
        \hline
    \end{tabular}
\end{table}

\section{\label{sc:Density} Density calculation}
\label{sec:galover}
\subsection{Galaxy density measurements}

The presence of different densities in the spatial distribution of galaxies reflects the intricate large-scale structure of the Universe. To measure projected local galaxy densities (hereafter densities), we use two methods: the aperture density; and the
$N$th-nearest-neighbour methods. Both quantify the local galaxy environment. 

The first
method consists in measuring the number of galaxies around a given galaxy within a fixed circular aperture.
The aperture density is defined as
\begin{equation}
\Sigma_{(r<R)} = \frac{N_{\rm gal} - N_{\rm bkg}}{N_{\rm bkg}} \, ,
\end{equation}
where $N_{\rm{gal}}$ is the number of galaxies within an aperture of radius $R$ from a given galaxy, and $N_{\rm{bkg}}$ is the mean number of background galaxies in the field within the same aperture. We adopt $R=$~\ang{;0.5;} and $R=$~\ang{;1;}, which correspond  approximately to 0.25 and 0.5 physical Mpc, respectively,  in our redshift range, $z>1.3$. These values are consistent with the aperture densities calculated by \cite{Wylezalek14} and \citet{Mei23}. 

The associated \gls{snr} is
\begin{equation}
\label{eq:snr}
    \mathrm{\gls{snr}}_{r<R} = \frac{N_{\rm gal} - N_{\rm bkg}}{\sigma_{\rm bkg}} \, ,
\end{equation}
where $\sigma_{\rm bkg}$ is the standard deviation of the background. The \gls{snr} provides a robust measure of the relative enhancement in galaxy density compared to background variations. Following \citet{Wylezalek14}, we calculate $N_{\rm bkg}$ and $\sigma_{\rm bkg}$ from the \gls{spuds}.  To compare with the literature, the figures in this paper use the surface density, which is defined as the number $N_{r<R}$ of selected galaxies within a circle of radius $R$, and does not depend on background estimates. This quantity is directly proportional to $\Sigma_{(r<R)}$. 
We calculate the average and standard deviation of the surface density distributions by applying a 3~$\sigma$ iterative clip, which discards densities at $>3~\sigma$ when deriving the average density and its standard deviation, until convergence.

Following \citet{Rettura18}, we estimate the foreground star contamination using the \citet{Wainscoat92} model,\footnote{\url{http://irsa.ipac.caltech.edu/applications/BackgroundModel/}} which predicts the number of optical-to-infrared point sources at a given position in the sky. For \gls{edff}, \gls{edfn}, and \gls{edfs} we find an average contamination of 0.72, 1.9, and 0.9 stars per arcmin$^2$, respectively. We correct our density measurement accordingly.

 
The second approach is the $N$th-nearest-neighbour method, which extends the analysis to larger spatial scales by calculating
the distance to the $N$th nearest-neighbour galaxy. It offers a broader perspective on galaxy density across varying scales. The density calculated with this method is defined as
\begin{equation}
\Sigma_N = \frac{N}{\pi D_N^2}\, ,
\end{equation}
where $N$ is the number of neighbouring galaxies and $D_N$ is the physical distance to the $N$th nearest neighbour in Mpc. 
\citet{Postman05} and \citet{Mei23} found consistent results for $N$ in the range of 5 to 10. For this paper, we adopt $N=7$, following  previous density estimates in clusters observed with the \acrlong{hst} \citep[\acrshort{hst};][]{Postman05} and in CARLA \citep{Mei23}.

 \begin{table}[!t]
    \caption{Aperture density ($\Sigma_{(r<\ang{;1;}}$) mean and standard deviation for the \gls{edf} and \gls{gaea} at the depth of the the S1 and S2 samples.}
    \label{tab:meanst}
    \centering
    \setlength{\tabcolsep}{8pt} 
    \begin{tabular}{lccc}
        \hline\hline\T\B
        Field&Sample&Mean & Standard deviation  \\
        \hline
        \noalign{\vskip 2pt}
       \gls{edfn}&S1  & 7.1 &2.3  \\
       & S2  &   10.0 & 2.8   \\
       \gls{edfn}&S1  & 7.0 & 2.3  \\
       & S2  &  9.3 & 2.8  \\
       EDF-S&S1  & 6.9 & 2.2 \\
       GAEA&  S2  &  9.8 & 3.3  \\
        \hline
    \end{tabular}
\end{table}

We applied these density measurements to samples of \spitzer-selected galaxies at $z>1.3$, and to passive-selected galaxies from \Euclid and ground-based $i$-band photometry from the \acrshort{dawn} survey. These galaxy samples are described in the next sections.

\subsection{\gls{irac}-selected sources}
\label{sec:ISS}
The galaxy spectral energy distribution presents an enhancement at $1.6\,\micron$ \citep{Sawicki02} as a prominent feature (bump) in the near-infrared spectra of galaxies, which is generated by $\mathrm{H}^-$ in the atmospheres of cool stars. 
With increasing redshift, this bump shifts progressively through the IRAC1 band and into the IRAC2 band. 
This spectral evolution causes galaxies in this redshift range to exhibit a higher flux in the IRAC2 band than in the IRAC1 band, resulting in a positive IRAC1$-$IRAC2 colour.
This phenomenon enables us to use a colour cut of ${\rm IRAC1} - {\rm IRAC2} > -0.1$ to select samples of massive galaxies at $z>1.3$ that are approximately $95\%$ complete and $95\%$ pure, regardless of their star formation activity and age \citep{Papovich08, Mei23}. 

Given the different magnitude limits of each \gls{edf} (see Table~\ref{tab:completeness}), for homogeneity we build two different galaxy samples to calculate galaxy densities. In both cases, we apply an upper magnitude limit of ${\rm IRAC2} > 18$ to exclude contamination from bright stars \citep{Wylezalek14}. Following \citet{Wylezalek14}, we also select only galaxies with a flux \gls{snr} of $ \mathrm{\gls{snr}}_\mathrm{f} > 3.5$ to keep only the most reliable detections. 

The first sample (hereafter S1) is selected at the magnitude limit of the shallowest field,  \gls{edfs}, which is ${\rm IRAC2} < 22.2$, at which all the \gls{edf} fields are at least 95\% complete. This selection ensures a homogeneous galaxy sample simultaneously in all three \gls{edf}.  This magnitude cut  corresponds approximately to a galaxy stellar mass of $\logten(M_*/M_{\odot}) \gtrsim 10.0 \pm 0.4$, following and adapting the calibration of \citet{Mei23}, which derives the galaxy mass from its correlation with IRAC1 magnitudes. 

The second sample (hereafter S2) is selected at deeper IRAC2 magnitude limits to compare it to density measurements available in the literature. We choose this sample magnitude limit to be similar to the CARLA cluster survey and the blank field survey from the \gls{spuds} \citep{galametz13} that have been used in the literature to select clusters and protoclusters at $z>1.3$ \citep{Wylezalek13, Wylezalek14}.

CARLA was carried out during \spitzer Cycles~7 and~8 using IRAC (P.I. D. Stern). CARLA focuses on detecting galaxy cluster candidates around \gls{rlqs} and \gls{hzrgs}, because radio sources are thought to trace dense regions of the Universe at high redshifts \citep[e.g.,][]{Hatch14,Daddi17}.  CARLA targets a sample of about 400 \gls{rlqs} and \gls{hzrgs} at $z > 1.3$. \citet{Wylezalek13} found that 46\% and 11\% of the CARLA densities are larger than those found in the SpUDS field at a 2$\sigma$ and a 5$\sigma$ level, respectively. About 20 of these CARLA overdense regions are spectroscopically confirmed and classified as clusters or protoclusters \citep{Noirot18}, and some of them show high local galaxy densities and high percentages of passive and early-type galaxies \citep{Noirot16, Mei23}. 

\gls{spuds} covers a field area of approximately 1\,deg$^2$ and was conducted as part of the \spitzer Cycle~4 Legacy Program. One of the primary goals of this survey was to study galaxy environments, including the influence of local and global surroundings on galaxy properties, such as star formation and quenching. For our density measurements in \gls{spuds}, we apply the same IRAC colour, magnitude, and \gls{snr} cuts as for the \gls{edf}.

Since the CARLA and \gls{spuds} fields are 95\% complete at ${\rm IRAC2}=22.9$, we select our second sample at a similar magnitude limit of ${\rm IRAC2}=22.8$, at which the \gls{edfn} is 95\% complete. For this comparison, we will consider only \gls{edfn} and \gls{edff}, which is 95\% complete at ${\rm IRAC2}=22.6$, while we exclude the shallower \gls{edfs}.  This magnitude cut corresponds approximately to a galaxy stellar mass of $\logten(M_*/M_{\odot}) \gtrsim 9.5 \pm 0.4$. Table~\ref{tab:bkg} gives the surface density and its standard deviation for our different samples and $R$ values.

\begin{figure}[t!]
\centering
\includegraphics[angle=0,width=1\hsize]{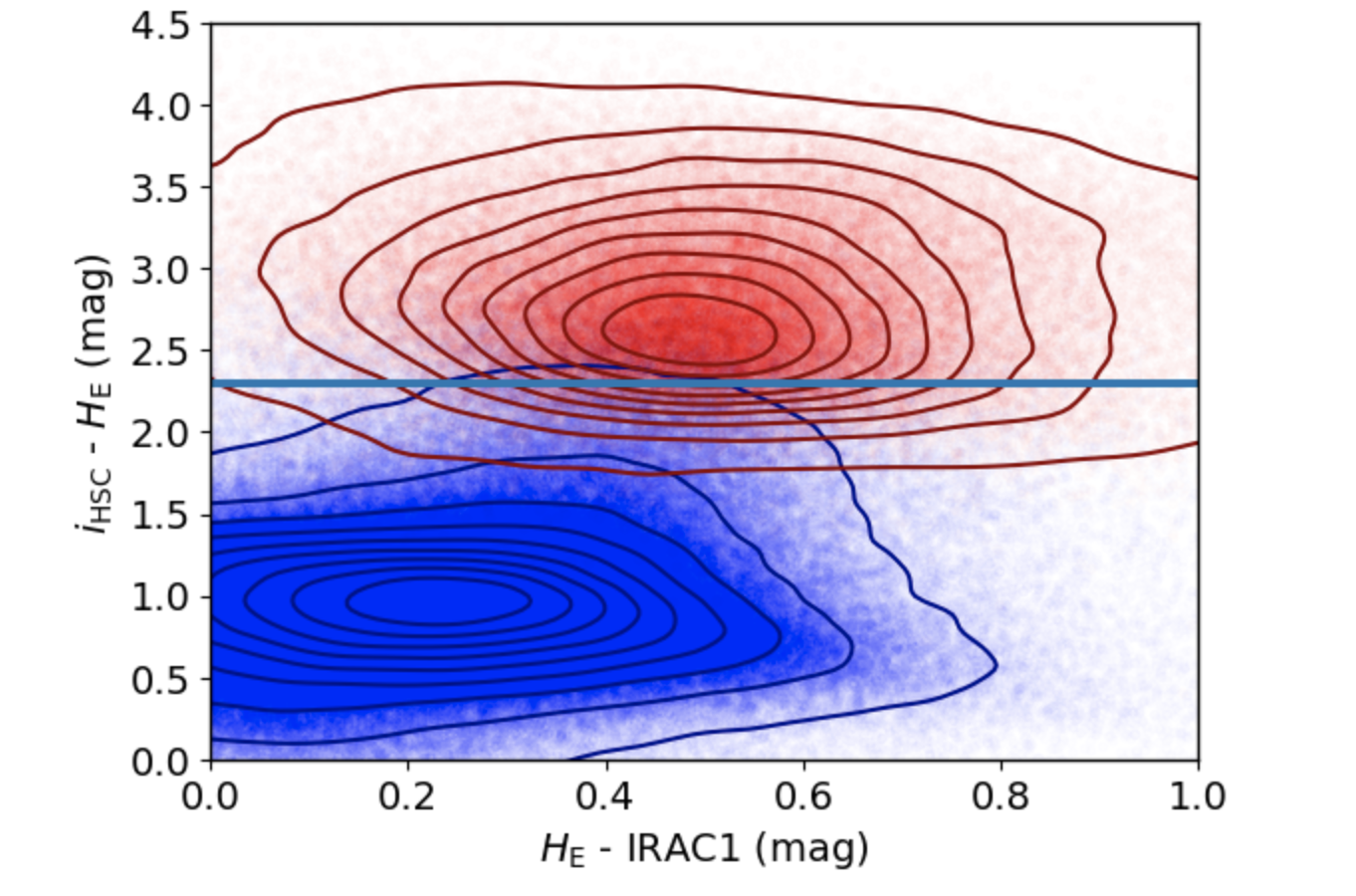}
\caption{Simulated observational (\ihsc - \HE) versus (\HE- IRAC1) colour-colour diagram for passive and star-forming galaxies from the \gls{gaea} simulations. Passive and star-forming galaxies are shown as red and blue points, respectively. The horizontal line  delineates the separation between passive and star-forming galaxies adopted in this paper, with passive galaxies cut at $(\ihsc - \HE) = 2.3$. This simple criteria is predicted to select passive galaxy samples that are about $90\% $ complete and $90\%$ pure. The contours represent the percentiles of the passive and star-forming galaxy distribution, in increments of 10\%. For example, the first and last contour from the centre of each distribution indicate that 10\% and 90\% of the galaxies, respectively, are within this contour.}
\label{fig:gaea}
\end{figure}

\subsection{Passive galaxy sample}
\label{sec:passive}

The goal of this paper is to establish the basis for combining \spitzer and \Euclid observations to extend the detection of clusters and protocluster candidates to redshifts $z>1.3$. To that end, we use \Euclid photometry from the \gls{q1} release and our \spitzer photometry and densities to identify extreme \spitzer overdense regions that host large numbers of passive galaxies. We aim to find interesting examples of potential cluster and protocluster candidates at $z>1.3$. A catalogue and classification of \gls{q1} clusters and protoclusters detected using our galaxy catalogues is beyond the scope of this paper. We will pursue the necessary in-depth density distribution analysis in a future work (Mei et al., in prep.).

These examples are interesting because \citet{Mei23} found that \spitzer-selected regions with high $\Sigma_{7}$ also show large fractions of passive galaxies out to $z\approx2$ \citep[see also ][]{Papovich08}. They demonstrated that apparent-magnitude colour-colour diagrams using ground-based $i$-band observations, \acrshort{hst} \gls{wfc3}, and \spitzer can separate passive from star-forming galaxies at $1<z<2$, in the same way as the widely used rest-frame \textit{UVJ} \citep{Williams09} and \textit{NUVJ} \citep{Arnouts13} diagrams. In fact, the \textit{i}-band, the \gls{wfc3} F160W (hereafter $H_{160}$), F140W (hereafter $H_{140}$), and the IRAC1 filters roughly correspond to rest-frame NUV, \textit{V}, and \textit{J}-band at these redshifts.
\citet{Mei23} demonstrated that these can be used to separate passive galaxies samples that are approximately $85\%$ complete and $85\%$ pure. A cut at high ($H_{160}-\mathrm{IRAC1}$) colour, typically at $(H_{160}-{\rm IRAC1}) \lesssim 1.6$, separates red passive galaxies from red dusty star-forming galaxies ($\lesssim 10\%$ of the red galaxies).

In \gls{edfn}, \gls{q1} \HE photometry, combined with deep \ihsc from the \acrshort{dawn} survey and IRAC1, provide apparent magnitudes that correspond to rest-frame NUV, \textit{V} and \textit{J}-band at $1.3 \lesssim z \lesssim 3$. To delineate the passive galaxy region in the ($\ihsc - \HE$) versus ($\HE- {\rm IRAC1}$) colour-colour diagram, we use the \gls{gaea} simulations, downgraded to the \gls{q1}, \acrshort{dawn}, and our IRAC1 photometric uncertainties in these filters. We start from the \gls{gaea} photometric catalogue, which does not have uncertainties, and add the \HE and \ihsc photometric uncertainties from \gls{q1} and the \acrshort{dawn} survey, and our IRAC1 photometric uncertainties. We extract photometric uncertainties from the observational distributions in each band in steps of 0.05\,mag. In this observational mock, we define a galaxy as quenched if the \gls{gaea} $\text{sSFR}<0.3\, t_{\text{H},z}^{-1}$ \citep{Franx2008}, where $t_{\text{H},z}$ is the Hubble time. A galaxy is defined as star-forming if it has $\text{sSFR} \ge 0.3\, t_{\text{H},z}^{-1}$.

\Cref{fig:gaea} shows the distribution of the simulated observational ($i_{\rm HSC}$ - \HE) versus (\HE- IRAC1) colour-colour diagram for passive and star-forming galaxies selected as above. The \gls{gaea} galaxy population, which does not manifest dusty star-forming galaxies, and the large uncertainties in the current \gls{q1} and \acrshort{dawn} catalogues do not permit us to define the equivalent of the \citet{Williams09} passive galaxy regions, as in \citet{Mei23}. In fact, when optimising the standard equations to separate passive from star-forming galaxies,
\begin{eqnarray*}
( \ihsc - \HE) &>& y_0 \, ,\\
(\HE- {\rm IRAC1}) &<&x_0\, , \\
(\ihsc - \HE) &>& a+b\times (\HE- {\rm IRAC1})\, ,
\end{eqnarray*}
we find that the only the first equation really matters, which efficiently separates passive from star-forming galaxies without further cuts. With $y_0=2.3$, we obtain a sample of passive galaxies that is approximately $90\%$ complete and $90\%$ pure. 

Given that \gls{gaea} reproduces observations of quenched galaxies in the local Universe very well \citep{Delucia24,Q1-SP017}, we apply this cut found with simulations, $(\ihsc - \HE) > 2.3 $, to observed \gls{q1} and \acrshort{dawn} apparent magnitudes to select a sample of passive galaxies that we predict to be $90\%$ complete and $90\%$ pure based on the GAEA simulations. We then calculate overdense regions of passive galaxies using the same methods described in \cref{sec:galover}. The average and standard deviation of the passive galaxy surface density distribution with ($R=\ang{;1;}$) are $\Sigma_{r<1\arcmin}=0.8\pm0.9$ at the depth of S2.

\section{Results}
\label{sec:results}
\subsection{Density distributions}
We present our density measurement distributions and compare them to those published for two other relevant \textit{Spitzer}-based surveys available in the literature: the \gls{spuds} blank field and the CARLA cluster survey. We also compare our results to predictions from the \gls{gaea} semi-analytical model \citep{Delucia24}. Since the published densities are surface densities with $R=\ang{;1;}$,  we focus only on surface density measurements with this aperture in this section. We recall that these are densities measured around each selected galaxy, and several might belong to the same structure; however, we do not perform structure detection in this paper. 

\begin{figure}[!t]
    \centering
    \includegraphics[width=0.8\hsize]{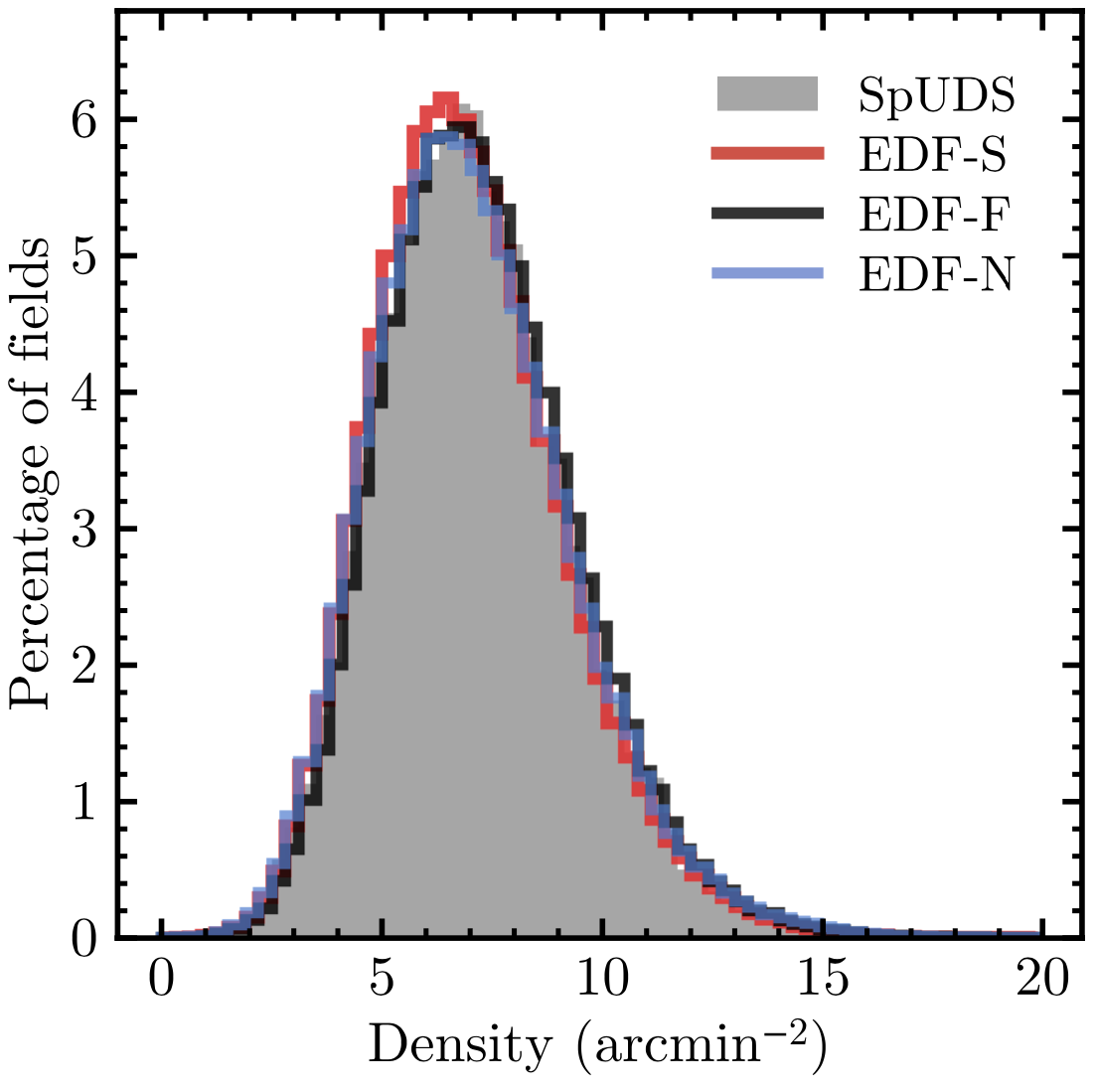}
    \caption{\spitzer surface density distributions ($R=\ang{;1;}$) for the three \gls{edf} and \gls{spuds}. We calculate surface densities at the same magnitude limit, ${\rm IRAC2}=22.2$, where all samples are at least 95\% complete. The four distributions show similar means and standard deviations (see \cref{tab:bkg,tab:meanst}). The distributions in the three \gls{edf} and in \gls{spuds} are consistent, and typical of blank fields. Only around $1\% $ of the surface densities measured in \gls{edfn} and \gls{edff}, and 0.003\% in \gls{edfs} are in overdense regions that are more than $3\sigma$ above the \gls{spuds} average surface density.}
    \label{fig:EDFs_density}
\end{figure}

\Cref{fig:EDFs_density} shows the S1 surface density distributions for the three \gls{edf} compared to \gls{spuds}.  At the same magnitude limit of ${\rm IRAC2}=22.2$, all samples are homogeneous and at least 95\% complete. \Cref{tab:meanst} lists the Gaussian mean and standard deviation of each distribution. This selection corresponds to galaxies with $\logten(M_*/M_{\odot}) \gtrsim 10.0 \pm 0.4$ (see \cref{sec:ISS}). 
For this massive galaxy sample, the density distributions in the three \gls{edf} and in \gls{spuds} are consistent, and typical of blank fields. Only about 1\% (1489 and 1499, respectively) of the densities measured in \gls{edfn} and \gls{edff}, and 0.003\% (1399) in \gls{edfs} are in overdense regions that are more than $3~\sigma$ above the \gls{spuds} average density. 

\begin{figure*}[!t]
    \centering
    \includegraphics[width=0.82\linewidth]{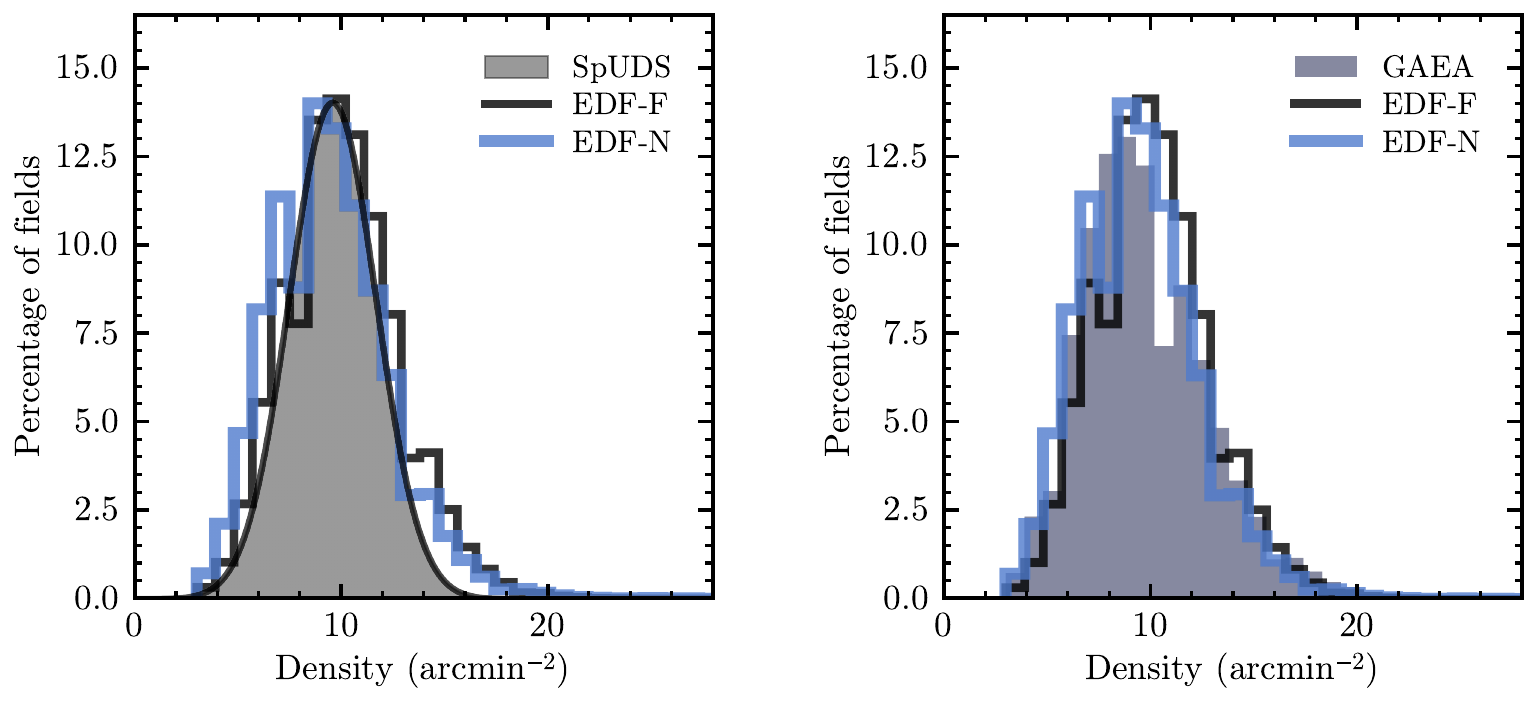}
    \caption{\spitzer surface density ($R=\ang{;1;}$) distributions in \gls{edfn} and \gls{edff} for our S2 samples, compared to the \gls{spuds} blank field and the \gls{gaea} simulations. The \gls{spuds} distribution is shown as a Gaussian. We find that the \gls{edfn} and \gls{edff}  have similar distributions and are consistent with predictions from the \gls{gaea} simulations. They, however, have a tail with 2--3\% of the galaxies at densities $3\sigma$ higher than the mean.}
    \label{fig:histodens}
\end{figure*}

However, published results of \spitzer surface densities reach an IRAC2 magnitude limit more similar to our S2 sample \citep[e.g., ][]{Rettura14, Wylezalek14, Martinache18, Mei23} than our S1 sample. We compare our S2 surface density measurements within the same radius ($R=\ang{;1;}$) to those that are publicly available for the CARLA and \gls{spuds} surveys \citep{Wylezalek14}. The deeper magnitude limit in S2 corresponds to galaxies with $\logten(M_*/M_{\odot}) \gtrsim 9.5 \pm 0.4$ (see \cref{sec:ISS}).

\Cref{fig:histodens} shows the surface density distributions of our S2 sample in the \gls{edfn} and \gls{edff} compared to those that we obtain in \gls{spuds} and in the \gls{gaea} simulations when applying the same galaxy cuts. \Cref{tab:meanst} gives the Gaussian mean and standard deviation of each distribution. The average density distributions and standard deviation of the \gls{edfn} and \gls{edff} fields are consistent between themselves. 

Interestingly, at this deeper magnitude and lower mass limit, 2\% (9048) and 3\% (9994) of the densities measured in the \gls{edfn} and \gls{edff} are at $3~\sigma$ from the \gls{spuds} average density. This is a clear indication that a large part of the less massive galaxies in the two \gls{edf} have larger densities than galaxies with the same mass in the field: there is high potential that they belong to groups or clusters \citep{Wylezalek14}. 

\begin{figure}[!t]
    \centering
    \includegraphics[width=0.8\hsize]{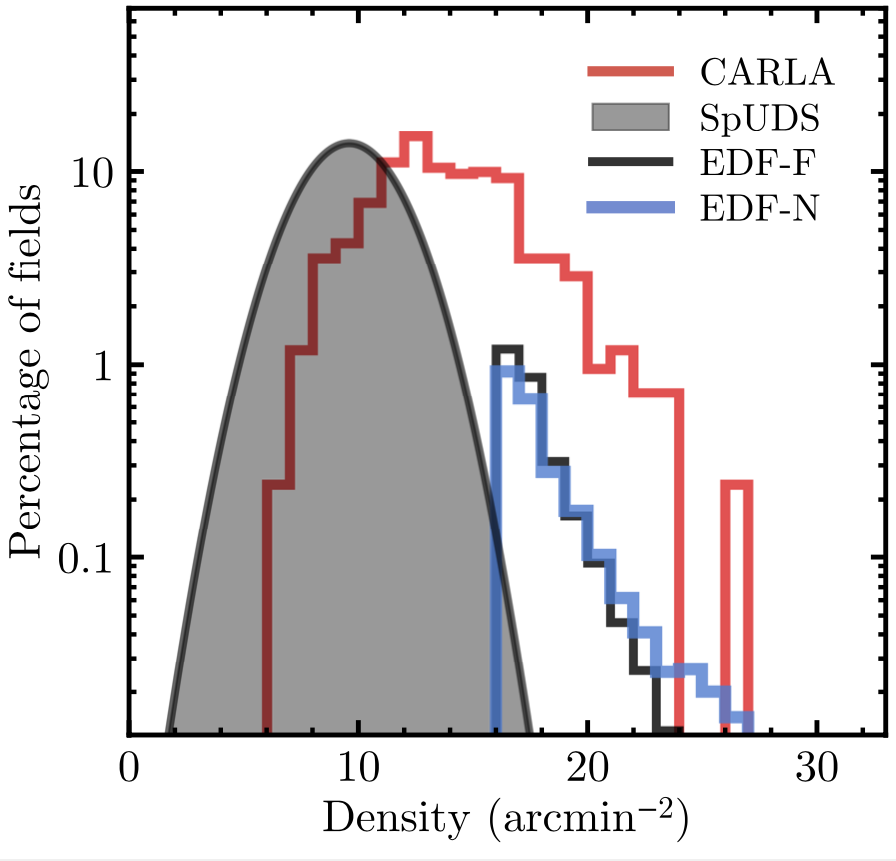}
    \caption{\spitzer surface density ($R=\ang{;1;}$) distributions for the \gls{edfn} and \gls{edff} S2 sample at $>3\sigma$ from the field mean, compared to \gls{spuds} and CARLA. This figure demonstrates that the \gls{edfn} and \gls{edff} show promising high surface densities consistent with those found in CARLA galaxy clusters. }
    \label{fig:threesig}
\end{figure}

To further illustrate this point, \cref{fig:threesig} shows the \gls{edf} surface densities at $ > 3\sigma$ from the \gls{spuds} distribution for the \gls{edfn} and \gls{edff}. These surface densities are comparable to the surface densities observed around the radio-loud AGN at the centres of CARLA clusters. This suggests that our measurements hold promise for detecting cluster and protocluster candidates in these fields, as \spitzer-selected overdense regions with respect to the field \citep{Wylezalek14}.

To compare with measurements of surface densities in the \gls{gaea} simulation, we degraded the \gls{gaea} magnitudes and added uncertainties as described in \cref{sec:passive}, using as reference the S2 \gls{edfn} \spitzer photometry and photometric uncertainties. We then measure densities in the same way as done for the observations. \gls{gaea} predictions for the density distribution are consistent with our results. 

 We note that the \gls{gaea} cosmology \citep{Springel05} differs from the cosmology adopted in this paper for the $\Sigma_7$ measurements. However, when calculating surface densities we do not use our adopted cosmological parameters, and we do not expect the different \gls{gaea} cosmology to substantially impact our results. To test this hypothesis, we compare measurements of $\Sigma_7$ in the degraded \gls{gaea} simulations, which do use our adopted cosmological parameters, for two different cosmologies -- the \citet{Springel05} cosmology and the \Planck 2015 cosmology. 
 
\Cref{fig:gaea_cosmo} shows the histogram of $\Sigma_7$ measurements with our chosen cosmology compared to the GAEA cosmology. This comparison demonstrates that the $\Sigma_7$ distributions, assuming the two different cosmologies, are consistent, and therefore we do not expect inconsistencies between our observed aperture measurements and those from GAEA, where cosmology affects in the same way only the galaxy distributions.

\subsection{Examples of the highest density regions}

As examples illustrating the potential of our measurements, we show the number of passive galaxies found in the 14 highest \spitzer-selected surface densities in \Cref{table:spitzer_overdensities} for the \gls{edfn} and \gls{edff}. These overdense regions are selected as having \gls{snr}$ > 20$ when using all three density measurements. When several densities satisfy this, we keep only the largest density within \ang{;2;}. We find that all of the highest-aperture densities in \gls{edfn} have $\Sigma^\mathrm{pass}_{r>1\arcmin} > 3.5~\sigma$. 

We also selected the 10 passive galaxies that have surface densities $> 10~\sigma$ from the average when using all three density measurements, which are therefore overdense regions of passive galaxies. Again, when several densities satisfy this, we keep only the largest density within \ang{;2;}. \Cref{table:passive} shows the passive-galaxy overdense regions and their \spitzer-selected densities. One of our passive galaxy highest densities, \gls{edfn} ID~1, corresponds to our \spitzer-delected highest density \gls{edfn} ID~4.


The presence of passive-galaxy overdense regions and of passive galaxies in \spitzer-selected overdense regions is consistent with large fractions of passive galaxies observed in the high-density regions of CARLA clusters \citep{Mei23}. This is very promising for the future use of our catalogues in the detection of clusters and protoclusters in the \gls{edf}.

We use the \gls{q1} photometric redshifts to estimate the redshift of these extreme densities, with the results given in \cref{table:spitzer_overdensities,table:passive}. However, \gls{q1} photometric redshifts have large uncertainties at $z>1$, and these redshift estimations should be taken with caution. For this reason, we only take galaxies with \gls{q1} photometric redshift uncertainties below 0.5, and we give an estimate of the overdensity average photometric redshift only when we have at least a $3~\sigma$ peak at a given redshift in the photometric redshift distribution within a circle of \ang{;1;} around our highest \spitzer-selected density. The uncertainties in our estimated photometric redshifts are given as the standard deviation of the photometric redshifts within $3~\sigma$ of the peak redshift in the region. These are very approximate estimations that we will refine when future, more precise, \Euclid photometric and spectroscopic data releases become available.
\Cref{fig:examples} shows images of the passive-galaxy highest densities with photometric redshift measurements.
 

\section{Cross-correlation with other galaxy ovedense regions in the \gls{edf}}
\label{sec:discussion}
We cross-correlate our \spitzer densities with cluster detections from the \gls{q1} cluster catalogue presented in Bhargava et al. (in prep.) at redshifts $z>1.2$, to take into account Q1 photometric redshift uncertainties.
We select \spitzer densities within \ang{;2;} of the B25 cluster positions and with $\Sigma_{r<1\arcmin}$ at $>3\sigma$ more than the field average. In \gls{edfs}, we use the S1 catalogue, while we use the S2 catalogues in \gls{edfn} and \gls{edff}. 
All cross-identifications are B25 clusters at $z>1.3$ within the cluster and density photometric redshift uncertainties, except cluster one which has a redshift $z=1.26 \pm 0.06$. 
In \gls{edfs}, \gls{edfn}, and \gls{edff}, we find that approximately 50\%, 30\%, and 10\%, respectively, of B25 clusters at $z>1.3$ (within the uncertainties) present \spitzer-selected galaxies densities more than $3\sigma$ above the field average. 

We also cross-correlate our densities with the  \acrlong{erosita} \citep[\acrshort{erosita},][]{Bulbul2024,Kluge2024} X-ray cluster catalogue. We do not find eRASS1 clusters at the position of our aperture density measurements that are $>3\sigma$ with respect to \gls{spuds}. However, this is expected, since we select galaxies at $z>1.3$, which is higher redshift than the eRASS1 cluster catalogue. It also confirms that our selection is not contaminated by lower redshift massive structures.


\section{\label{sc:Conclusions} Conclusions}
We have combined \Euclid and \spitzer observations of the \gls{q1} \gls{edf} as a basis for extending the detection of \Euclid clusters and protoclusters to $z>1.3$. We measured \spitzer-selected galaxy densities at $z>1.3$ and found that 2--3\% of the surface densities measured in the \gls{edfn} and \gls{edff} fall $3\sigma$ above the mean density; these overdense regions are consistent with galaxy densities measured in the CARLA cluster sample at $z>1.3$. 

We also find that the \spitzer-selected overdense regions exhibit overdense regions of passive galaxies selected with combined \Euclid and ancillary \acrshort{dawn} observations. These results confirm the promise of our catalogues for detecting cluster and protocluster candidates in these fields, as \spitzer-selected overdense regions with respect to the field \citep{Wylezalek14}. 

We built a catalogue of \spitzer-selected galaxy densities at redshift $z > 1.3$ in the \gls{edf} from archival data  \citep{Moneti22}. Source detection was performed with the \sextractor software, and bright sources were masked to prevent contamination. Our catalogue was validated with the COSMOS2020 catalogue, a data set with a deeper completeness magnitude limit. Using this latter catalogue as a complete reference sample, we found approximately 95\% completeness at magnitude limits of ${\rm IRAC1}=22.9$ and ${\rm IRAC1}=23.2$ for \gls{edfn} and \gls{edfn}, respectively, and at ${\rm IRAC1}=22.4$ for the \gls{edfs}. 

To measure galaxy densities, we used two complementary methods: (1) aperture and surface density measurements at two different aperture radii ($R=\ang{;1;}$ and \ang{;0.5;}); and (2) the \textit{N}th-nearest-neighbour method.  
Given that \spitzer observations in the \Euclid \gls{q1} fields have different depths, we built two galaxy samples. 
\begin{itemize}
    \item S1, with a magnitude limit of ${\rm IRAC2} < 22.2$, at which all \gls{edf} are at least 95\% complete; this magnitude limit corresponds to galaxies with $\logten(M_*/M_{\odot}) \gtrsim 10.0 \pm 0.4$. 
    
\item S2, at the depth of  ${\rm IRAC2} < 22.8$, at which the \gls{edfn} is $95\%$ complete, and which permits us to compare our results to other results already published in the literature; specifically, the CARLA cluster and \gls{spuds} field surveys. This magnitude limit corresponds to galaxies with $\logten(M_*/M_{\odot}) \gtrsim 9.5 \pm 0.4$.
\end{itemize}

For the most massive galaxy sample, S1, the density distributions in the three \gls{edf} and in \gls{spuds} are consistent, and typical of blank fields. In fact, only about 1\%  of the densities measured in \gls{edfn} and \gls{edff}, and 0.003\% in \gls{edfs}, are more than $3~\sigma$ above the \gls{spuds} average density. 

However, when considering the deeper magnitude limit sample, S2, 2\% and 3\% of the densities measured in the \gls{edfn} and \gls{edff}, respectively, are at more than $3~\sigma$ from the \gls{spuds} average density. These overdense regions are characteristic of densities found in CARLA clusters. This is a clear indication that a large part of the less massive galaxies in the two \gls{edf} have larger densities than galaxies with the same mass in the field: there is high potential that they belong to groups or clusters \citep{Wylezalek14}. We plan a more detailed analysis for cluster and protocluster detection in future work (Mei et al., in prep.). 

This result was further confirmed when we measured passive galaxy densities and found passive-galaxy overdense regions in \spitzer-selected overdenties. In fact, \spitzer-selected clusters and protoclusters in the literature show a high fraction of passive galaxies in their high-density regions out to $z\approx 2$ \citep{Mei23}.

Our \spitzer photometry and density catalogues will be made available upon publication. \Cref{table:spitzer_photometry} presents the structure of these final catalogues.

%
%

\begin{acknowledgements}
  This work was supported by CNES, focused on the Euclid space mission. NM has received support from France 2030 through the project named Académie Spatiale d'Île-de-France (\url{https://academiespatiale.fr/}) managed by the National Research Agency under bearing the reference ANR-23-CMAS-0041, and from CNRS/IN2P3. Numerical computations were partly performed on the DANTE platform, APC, France. Based on data from UNIONS, a scientific collaboration using three Hawaii-based telescopes: CFHT, Pan-STARRS, and Subaru
\url{www.skysurvey.cc}\,. Based on
data from the Dark Energy Camera (DECam) on the Blanco 4-m Telescope
at CTIO in Chile \url{https://www.darkenergysurvey.org}\,.
This work is based on observations made with the Spitzer Space Telescope, which was operated by the Jet Propulsion Laboratory, California Institute of Technology under a contract with NASA. Support for this work was provided by NASA through an award issued by JPL/Caltech\,.
This work has made use of the Euclid Quick Release Q1 data
from the Euclid mission of the European Space Agency (ESA), 2025, https:
//doi.org/10.57780/esa-2853f3b\,.
\AckERO 
\AckEC 
\AckCosmoHub
\end{acknowledgements}

%
%

\bibliography{Q1}

%
%
\begin{appendix}

\section{Figures}

\begin{figure*}[hp]
    \centering
    \includegraphics[width=0.4\linewidth]{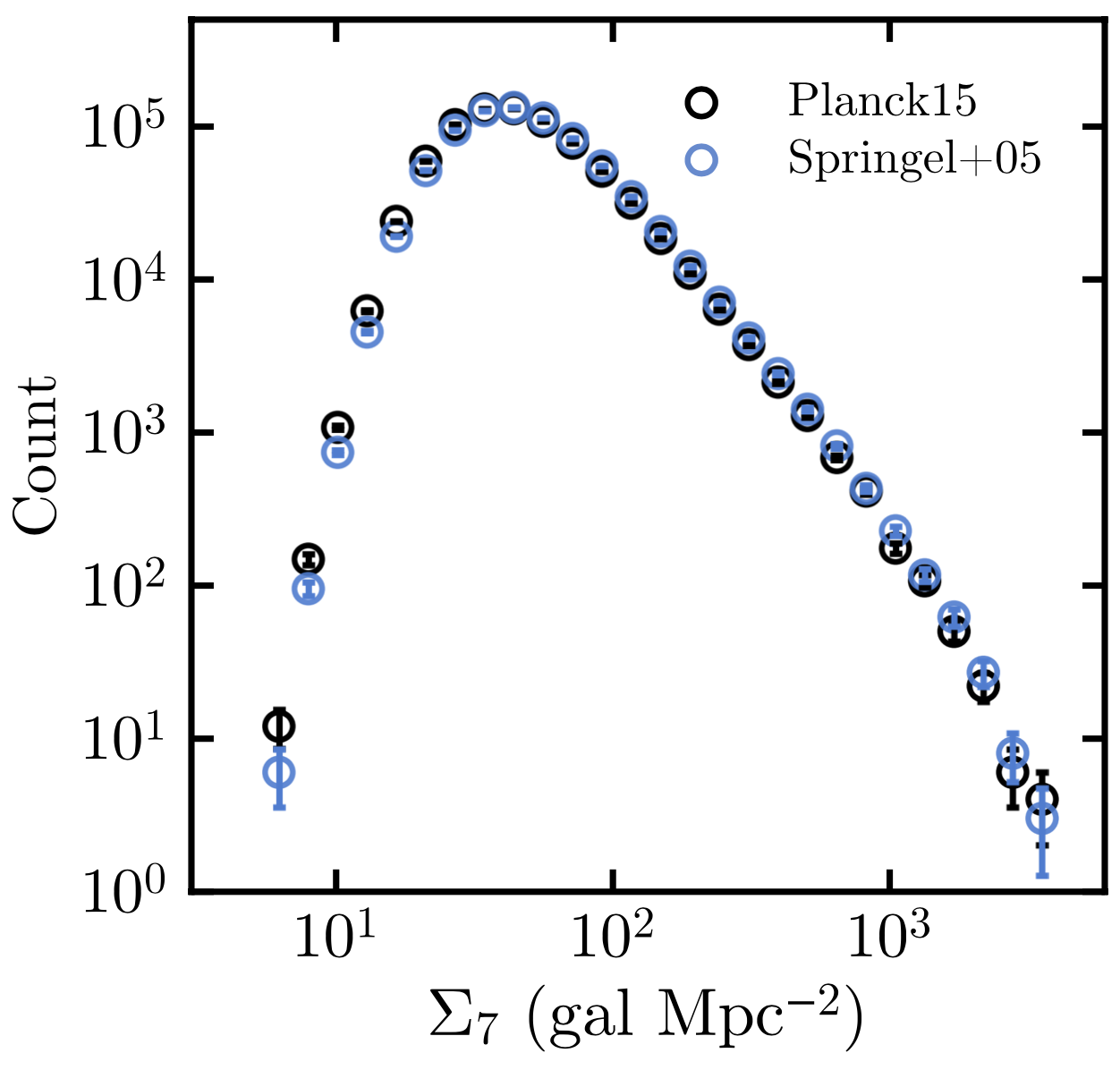}
    \caption{Number count distribution of $\Sigma_7$ densities in \gls{gaea} for both the \citet{Planck15} cosmology and the \textit{Millennium} simulation cosmology \citep{Springel05}. Poisson errors are included as error bars. The two distributions are consistent within their mutual uncertainties, indicating that the choice of cosmology should not impact our results.}
    \label{fig:gaea_cosmo}
\end{figure*}

\Cref{fig:gaea_cosmo} shows the number count distribution of $\Sigma_7$ densities in the \gls{gaea} simulations for two different cosmologies: the \citet{Planck15} cosmology and the \textit{Millennium} simulation cosmology \citep{Springel05}. The two distributions are consistent. This means that the choice of cosmology should not impact our results.

\begin{figure*}[h]
\center
\includegraphics[width=0.32\textwidth]{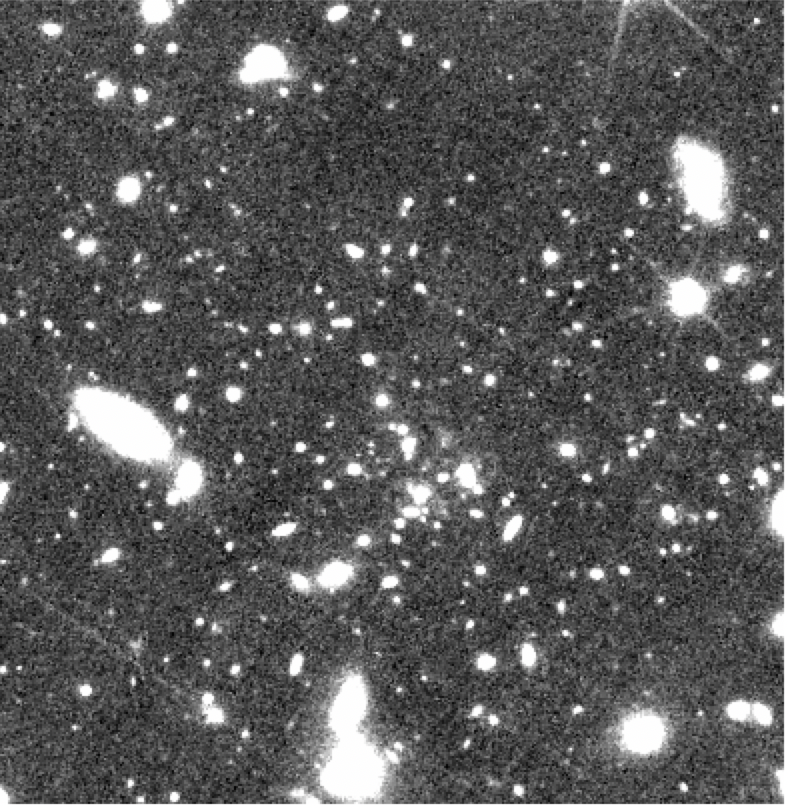}
\includegraphics[width=0.32\textwidth]{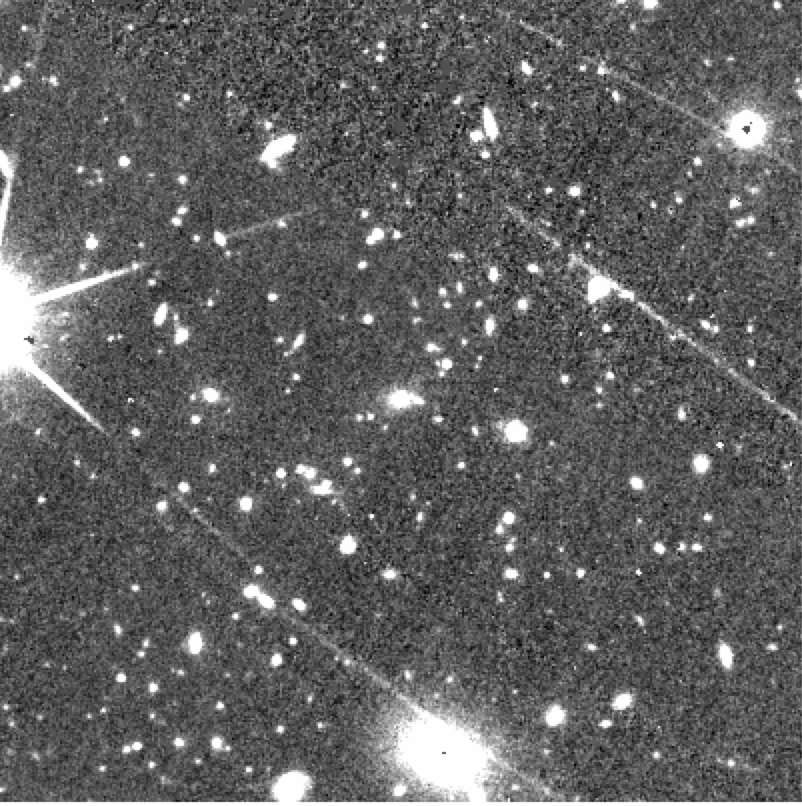}
\includegraphics[width=0.32\textwidth]{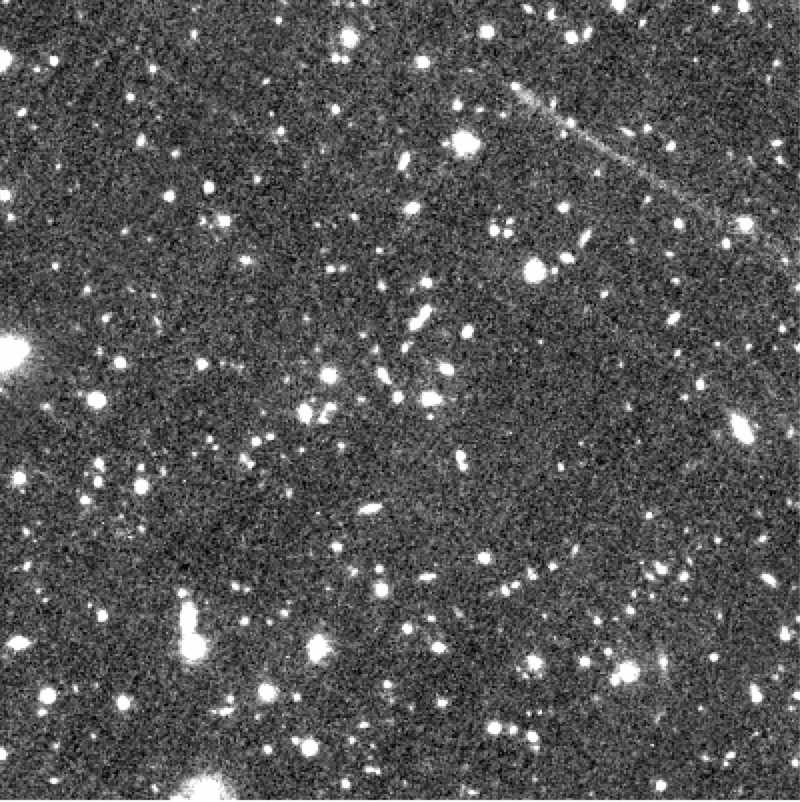}
\includegraphics[width=0.32\textwidth]{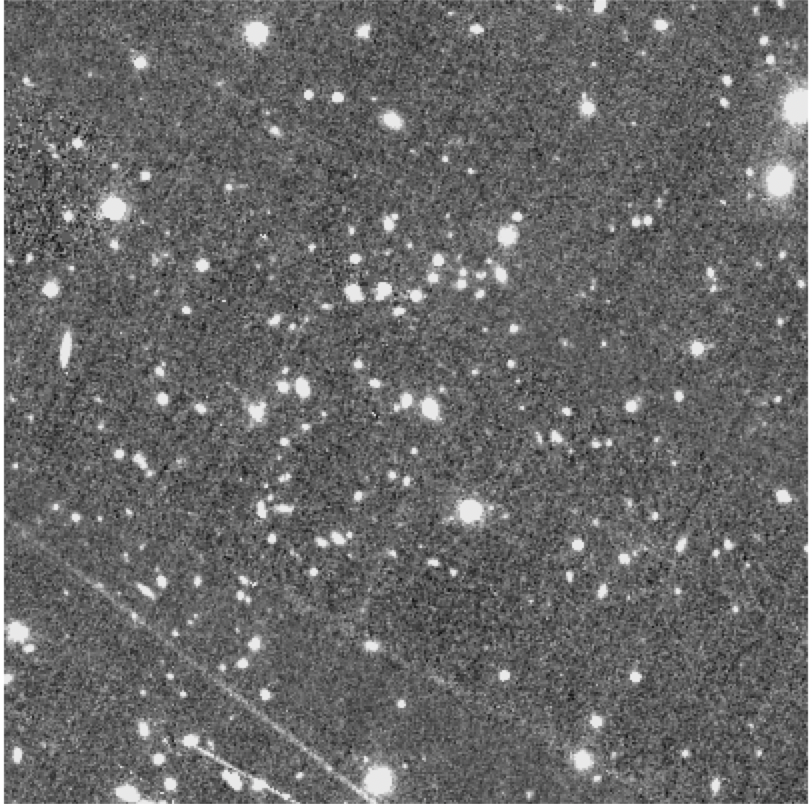}
\includegraphics[width=0.32\textwidth]{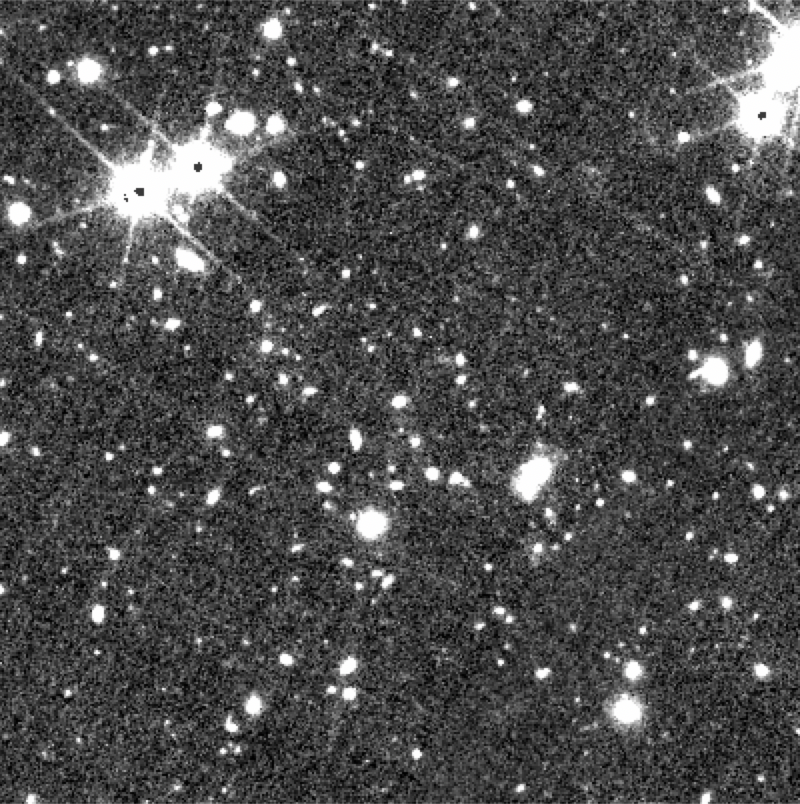}
\caption{Examples of our \spitzer-selected highest-density regions with high densities of passive galaxies. The size of each image is $2\arcmin \times 2\arcmin$. From left to right and top to bottom: density ID 1, 3, 9, 2, 4, at redshift $1.39 \pm 0.05$, $1.39 \pm 0.05$, $1.39 \pm 0.05$, $1.46 \pm 0.04$, and $1.54 \pm 0.06$, respectively, as from \cref{table:passive}.}\label{fig:examples}
\end{figure*}

\Cref{fig:examples} shows examples of our \spitzer-selected highest-density regions with high densities of passive galaxies, presented in \cref{table:passive}.

\newpage
\clearpage

\section{Tables}

\Cref{tab:sexpar} shows the \sextractor key parameters used for source detection when performing IRAC photometry.

\Cref{tab:tphotpar} shows the parameters used when performing IRAC photometry using \tphot \citep{Merlin15,Merlin16}. The \tphot pipeline consists of several steps and works the best when running two passes. The first pass (1pass) (1) creates multiwavelength stamps for the galaxies in the input catalogue of the high-resolution image (priors); (2) then convolves these stamps with a kernel that reduces the stamps to the low-resolution image (convolve); (3) performs the fitting procedure to model each source flux in the low-resolution image (fit); (4) chooses the best fit (diags); (5) computes a list of positional shifts, and a set of shifted kernels are generated and stored (dance).The second pass repeats steps (2), (3) and (4) above, and then archives the results (archive). In the priors stage, we can choose to use observed pixel values (usereal), which is what we do, or a model (usemodels) or indicate that we have unresolved sources (useunresolved). If the convolution is performed in the Fourier space, which is our case, the parameter FFTconv is set as true. In the fitting stage, we: (1) use the cells-on-object algorithm (coo), for which an optimized number of objects is considered in a cell to minimize bias; (2) activate as true the cellmask option, which excludes pixels from
the fit if they are below a given floor (maskfloor); (3) do not fit the background (fitbackground); (4) do not apply a threshold (threshold) that defines the pixels above it that are used in the fitting process; (5) choose the Cholesky solution method (lu); (6) and exclude negative solutions (clip). For details on each step, please refer to \citet{Merlin15,Merlin16}.

\Cref{table:spitzer_overdensities} shows examples of our S2 \spitzer-selected highest-density regions.  \Cref{table:passive} presents the ten largest passive-galaxy densities in \gls{edfn}. \Cref{table:spitzer_photometry} shows the structure of our catalogues of \spitzer photometry and projected galaxy densities. 

\begin{table}[hb]
    \caption{\sextractor key parameters used for source detection in all fields.}
    \label{tab:sexpar}
    \centering
    \setlength{\tabcolsep}{10pt} 
    \begin{tabular}{lc}
        \hline\hline\T\B
        Parameter & Value \\
        \hline
        DETECT\_MINAREA  & 3.0    \T \\
        DETECT\_THRESH    & 1.8     \\
        ANALYSIS\_THRESH  & 1.8     \\
        DEBLEND\_NTHRESH  & 32      \\
        DEBLEND\_MINCONT  & 0.0001  \\
        BACK\_SIZE        & 16      \\
        BACK\_FILTERSIZE  & 3       \\
        BACKPHOTO\_THICK  & 32      \\
        BACKPHOTO\_TYPE   & LOCAL  \B \\
        \hline
    \end{tabular}
\end{table}

\begin{table*}[hp]
\caption{\tphot parameters.}
\label{tab:tphotpar}
\centering
\begin{tabular}{lcccccc}
\hline\hline 
\multirow{2}{*}{Pipeline} & 1st pass & priors, convolve, fit, diags, dance \\
& 2nd pass & convolve, fit, diags, archive \\ 
\hline 
\multirow{3}{*}{Priors stage} & usereal & true \\
& usemodels & false \\
& useunresolved & false \\ 
\hline 
\noalign{\vskip 2pt}
Convolution stage & FFTconv & true \\ 
\hline 
\multirow{7}{*}{Fitting stage} & fitting & coo \\
& cellmask & true \\
& maskfloor & $10^{-9}$ \\
& fitbackground & false \\
& threshold & 0.0 \\
& linsyssolver & lu \\
& clip & true \\
\hline
\end{tabular}
\end{table*}

\begin{table*}[hp]
    \caption{Examples of our S2 \spitzer-selected highest-density regions. These have been selected as having densities $>20 \ \sigma$ with all our three methods.}
    \label{table:spitzer_overdensities}
       \begin{tabular}{lccccccccccccc}
        \hline 
        \rule{0pt}{2.5ex}Q1 Field & ID&RA & Dec & $z_{\rm phot}$ & $N_{\rm zphot}$  & $N^{\rm pass}{(r<1\arcmin)}$ &  $\Sigma_{(r<1\arcmin)}$ & \gls{snr}$_{(r<1\arcmin)}$ & $\Sigma_7$ &  \gls{snr}$_{\Sigma_7}$ \\
        \hline
        \rule{0pt}{2.5ex} \gls{edfn} & 1&272.13150 & $+67.17863$ &       & & 4 & 5 & 25 & 288 & 130 \\
        \rule{0pt}{2.5ex} \gls{edfn} &2& 268.59576 & $+64.89498$ &        & & 7 & 5 & 24 & 170 & 76 \\
        \rule{0pt}{2.5ex} \gls{edfn} &3& 270.68832 & $+65.35073$ &        & & 4 & 5 & 23 & 225 & 100 \\
        \rule{0pt}{2.5ex} \gls{edfn} &4& 266.79734 & $+65.54727$ & $1.39\pm0.03$ & 57 &  15 & 5 & 23 & 181 & 80 \\
        
        \rule{0pt}{2.5ex} \gls{edfn} &5& 270.82411 & $+65.06381$ & $1.36\pm0.02$ & 56  & 6 & 5 & 22 & 168 & 75 \\
        
        \rule{0pt}{2.5ex} \gls{edfn} & 6&270.11342 & $+66.25161$ & $1.46\pm0.02$ & 56  & 9 & 5 & 22 & 257 & 120 \\
        \rule{0pt}{2.5ex} \gls{edfn} &7& 272.93821 & $+67.03152$ &       &  & 8 & 5 & 23 & 161 & 70 \\
        \rule{0pt}{2.5ex} \gls{edfn} &8& 267.37057 & $+66.53104$ &       &  & 4 & 5 & 21& 332 & 150 \\
        \rule{0pt}{2.5ex} \gls{edfn} &9&269.59719 & $+64.65159$ &       &  & 6 & 5 & 21 & 157 & 70 \\
        \rule{0pt}{2.5ex} \gls{edfn} &10& 266.85880 & $+65.72311$ &       &  & 8 & 5 & 21 & 145 & 64 \\
        \rule{0pt}{2.5ex} \gls{edff} &11& 52.33491 & $-28.50047$ &        & &  & 6 & 25 & 216 & 100 \\
        \rule{0pt}{2.5ex} \gls{edff} &12& 51.60704 & $-27.17787$ &       & &  & 5 & 24 & 230 & 100 \\
        \rule{0pt}{2.5ex} \gls{edff} &13& 54.14683 & $-28.59636$ &        & &  & 5 & 22 & 192 & 90 \\
        \rule{0pt}{2.5ex} \gls{edff} &14& 51.92977 & $-27.67554$ &        & &  & 5 & 21 & 214 & 100 \\
        \hline
    \end{tabular}
    \begin{flushleft}
        \footnotesize\textbf{Notes.} The columns are: \gls{q1} field indicates the \gls{edf} hosting the densities; ID is the density ID; right ascension, RA, and declination, Dec, give the  position; $z_\mathrm{phot}$ is an estimate of the redshift from the \gls{q1} photometric redshift release (see text); $N_\mathrm{zphot}$ is the number of galaxies used for the photometric redshift estimate (see text); $N_{\rm pass}$ is the number of selected passive galaxies within a \ang{;1;} radius aperture from the \spitzer-selected density; $\Sigma_{r<R\arcmin}$ is the aperture density and  \gls{snr}$_{r<R\arcmin}$ is its S/N; $\Sigma_7$ is the density calculated with the $N$th-nearest-neighbour method in units of galaxies per Mpc$^2$ and \gls{snr}$_7$ is its S/N. We remind the reader that we can select passive galaxies only in \gls{edfn}, and this is why we don't find passive overdense regions in \gls{edff}.
    \end{flushleft}
\end{table*}

\begin{table*}[hp]
    \caption{The ten largest passive-galaxy densities in \gls{edfn}, with densities $>10 \ \sigma$  from the average.}
    \label{table:passive}
       \begin{tabular}{lcccccccccccc}
        \hline 
        \rule{0pt}{2.5ex} Q1 Field & ID&RA & Dec & $z_{\rm phot}$ &$N_{\rm zphot}$ &$N^{\rm pass}{(r<1\arcmin)}$ & $\Sigma^\mathrm{pass}_{(r<1\arcmin)}$ &  \gls{snr}$^\mathrm{pass}_{(r<1\arcmin)}$ & $\Sigma^\mathrm{pass}_7$ &  \gls{snr}$^\mathrm{pass}_{\Sigma_7}$ \rule[-1ex]{0pt}{0pt}\\
        \hline
        \rule{0pt}{2.5ex} \gls{edfn} & 1&266.80351 & $+65.55044$ &1.39 $\pm$ 0.05  & 15&15 & 6 & 13 & 6 $\pm$ 2& 30  \\
        
        \rule{0pt}{2.5ex} \gls{edfn} &2& 266.88792 & $+66.06646$  &1.46 $\pm$ 0.04 & 13 &13 & 5 & 11 & 4 $\pm$ 2& 20  \\
        
        \rule{0pt}{2.5ex} \gls{edfn} &3& 267.48010 & $+66.07398$ &  1.39 $\pm$ 0.05&12 &17 & 7 & 15 & 6 $\pm$ 2& 30  \\
        
        \rule{0pt}{2.5ex} \gls{edfn} &4& 267.68164 & $+67.30834$ & 1.54 $\pm$ 0.06 &11&  13 & 5 & 11 & 4 $\pm$ 2& 20 \\
        
        \rule{0pt}{2.5ex} \gls{edfn} &5& 267.85836 & $+66.36558$ & & &14 & 6 & 12 & 8 $\pm$ 3& 40\\
        
        \rule{0pt}{2.5ex} \gls{edfn} &6& 268.47776 & $+64.88699$ & &  &13 & 5 & 11 & 3 $\pm$ 1& 10  \\
        
        \rule{0pt}{2.5ex} \gls{edfn} &7& 268.79228 & $+67.69032$ &  &&15 & 6 & 13 & 9 $\pm$ 4 & 45  \\
        
        \rule{0pt}{2.5ex} \gls{edfn} &8& 269.39637 & $+65.22664$ &   &&15 & 6 & 13 & 3 $\pm$ 1 & 14 \\
        
        \rule{0pt}{2.5ex} \gls{edfn} & 9&271.82177 & $+65.49944$ & 1.39 $\pm$ 0.05& 12&13 & 5 & 11 & 3 $\pm$ 1 & 14  \\
        
        \rule{0pt}{2.5ex} \gls{edfn} &10& 271.99077 & $+65.71248$ &   &&13 & 5 & 11 & 4$\pm$ 2 & 20 \\
        
        \hline
    \end{tabular}
    \begin{flushleft}
        \footnotesize\textbf{Notes.} The columns are: \gls{q1} field indicates the \gls{edf} hosting densities; ; ID is the density ID; right ascension, RA, and declination, Dec, give the position; $z_\mathrm{phot}$ is an estimate of the redshift from the \gls{q1} photometric redshift release (see text); $N_\mathrm{zphot}$ is the number of galaxies used for the photometric redshift estimate (see text); $N_{\rm pass}$ is the number of selected passive galaxies within a 1\arcmin radius aperture from the \spitzer-selected density; $\Sigma^\mathrm{pass}_{r<R\arcmin}$ is the aperture density of passive galaxies and  SNR$^\mathrm{pass}_{r<R\arcmin}$ its S/N; $\Sigma^\mathrm{pass}_7$ is the passive galaxy density calculated with the $N$th-nearest-neighbour method in units of galaxies per Mpc$^2$ and  SNR$^\mathrm{pass}_7$ its S/N.
    \end{flushleft}
\end{table*}

\begin{table*}[hp]
    \caption{\spitzer photometry and projected galaxy densities: catalogue column structure} 
    \label{table:spitzer_photometry}
    \begin{tabular}{lcccccccccccccccc}
        \hline 
         \rule{0pt}{2.5ex} Q1 field & Cluster ID & RA & Dec & $z_\mathrm{phot}$  & $\Sigma^\mathrm{max}_7$  & $N^\mathrm{max}_{1\arcmin}$ &  \gls{snr}$^\mathrm{N}_{(r<1\arcmin)}$ & $N^\mathrm{max}_{(r<0.5\arcmin)}$ &  \gls{snr}$^\mathrm{N}_{(r<0.5\arcmin)}$  \rule[-1ex]{0pt}{0pt}\\\rule[-1ex]{0pt}{0pt}\\
        \hline
        \rule{0pt}{2.5ex} EDFN & $266.88946$ & $+66.06738$ & ... & .. & ..  & ... & ... & ... & ... \\
        \rule{0pt}{2.5ex} EDFN & $270.88027$ & $+65.72358$ & ... & .. & ..  & ... & ... & ... & ...  \\
        \rule{0pt}{2.5ex} EDFN & $267.36383$ & $+66.52386$ & ... & .. & ..  & ... & ... & ... & ...  \\
        \hline
    \end{tabular}
    \begin{flushleft}
        \footnotesize{\textbf{Notes.} The columns are: \gls{q1} field indicates the \gls{edf} hosting the densities; right ascension, RA, and declination, Dec, give the  position; $\Sigma_7$ is the density calculated with the $N$th-nearest-neighbour method in units of galaxies per Mpc$^2$ and  SNR$_7$ its S/N; $\Sigma_{r<R\arcmin}$ is the aperture density and SNR$_{r<R\arcmin}$ its S/N. } 
    \end{flushleft}
     \label{LastPage}
\end{table*}

\end{appendix}

\label{LastPage}
\end{document}